\documentclass[12pt, appendixfloats, numberedappendix]{emulateapj}

\usepackage{hyperref}
\usepackage{color}
\usepackage{natbib}
\usepackage{rotating}
\usepackage{amsmath,tabu}

\def\eg{e.g., }

\newcommand{\beq}{\begin{equation}}
\newcommand{\eeq}{\end{equation}}

\def\hyi{\ion{H}{1}}

\def\hytwo{H$_2$}

\newcommand{\kms}{\ensuremath{{\rm km~s}^{-1}}}

\newcommand{\ebv}{\rm \ensuremath{E_{B-V}}}
\newcommand{\Nhyi}{\rm \ensuremath{N_{\rm{HI}}}}
\newcommand{\Nhytwo}{\ensuremath{N_{\rm{H_2}}}}

\usepackage{threeparttable}

\begin{document}

\title{Exploring the diffuse interstellar bands with the Sloan Digital Sky Survey}
\shorttitle{Diffuse Interstellar Bands in SDSS spectra}

\shortauthors{Lan, M\'enard \& Zhu}
\author{
Ting-Wen Lan\altaffilmark{1},
Brice M{\'e}nard\altaffilmark{1,2,3} \&
Guangtun Zhu\altaffilmark{1,4}
} 
\altaffiltext{1}{Department of Physics \& Astronomy, Johns Hopkins University, 3400 N. Charles Street, Baltimore, MD 21218, USA, tlan@pha.jhu.edu}
\altaffiltext{2}{Kavli IPMU (WPI), the University of Tokyo, Kashiwa 277-8583, Japan}
\altaffiltext{3}{Alfred P. Sloan Fellow} 
\altaffiltext{4}{Hubble Fellow} 
\begin{abstract}
We use star, galaxy and quasar spectra taken by the Sloan Digital Sky Survey to map out the distribution of diffuse interstellar bands (DIBs) induced by the Milky Way. After carefully removing the intrinsic spectral energy distribution of each source, we show that by stacking thousands of spectra, it is possible to measure statistical flux fluctuations at the $10^{-3}$ level, detect more than 20 DIBs and measure their strength as a function of position on the sky. We create a map of DIB absorption covering about 5000 deg$^{2}$ and measure correlations with various tracers of the interstellar medium: atomic and molecular hydrogen, dust and polycyclic aromatic hydrocarbons (PAHs). After recovering known correlations, we show that each DIB has a different dependence on atomic and molecular hydrogen: while they are all positively correlated with \Nhyi, they exhibit a range of behaviours with \Nhytwo\ showing positive, negative or no correlation. We show that a simple parametrization involving only \Nhyi\ and \Nhytwo\ applied to all the DIBs is sufficient to reproduce a large collection of observational results reported in the literature: it allows us to naturally describe the relations between DIB strength and dust reddening (including the so-called skin effect), the related scatter, DIB pair-wise correlations and families, the affinity for $\sigma/\zeta$-type environments and other correlations related to molecules. Our approach allows us to characterize DIB dependencies in a simple manner and provides us with a metric to characterize the similarity between different DIBs.
\end{abstract}

%% =======================================
\section{Introduction}
%% =======================================

The diffuse interstellar bands (DIBs) are a set of absorption features observed ubiquitously in the interstellar medium (ISM). The first features, reported by \cite{Heger1922}, were identified as interstellar in origin by \cite{Merrill1934}. In the past eighty years, the list of DIBs has increased to more than 500 features \citep[][]{Hobbs2008,Hobbs2009}. However, despite active research, the identity of the DIB carriers remains unknown until this day. This has been one of the longest standing problems in astronomy \citep[for a review, see][]{Herbig1995}. Candidates of carriers include complex carbonaceous gas-phase molecules, such as fullerenes \citep{Foing1994} and polycyclic aromatic hydrocarbons \citep[PAHs, e.g.,][]{SalamaPAH1996}. However, none of them has yet been convincingly shown to be associated with any particular DIB.

\begin{figure*}[ht]
\begin{center}
\includegraphics[scale=0.4]{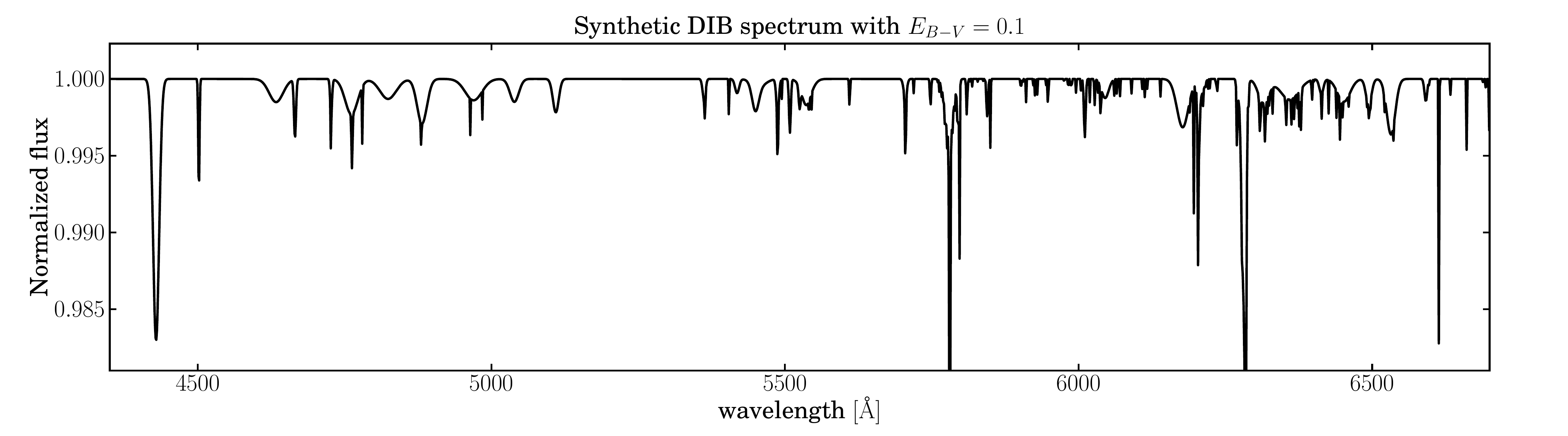}
\caption{A synthetic spectrum of DIB absorption (from the compilation by \citealt{Jenniskens1994}) representative of a line of sight with \ebv$\sim0.1$ mag. Note the expected absorption  is at the 1\% level.}
\label{plot:synthetic_spectrum}
\end{center}
\end{figure*}

DIBs are mostly found in the optical and near-infrared, with the longest reported wavelength at $1.793\,\mu$m \citep{Geballe2011}. They display a large range in width and central depth. The narrowest lines have full width at half-maximum (FWHM) less than $1\,$\AA\ while the broad DIBs have FWHM reaching $30\,$\AA. 
The central depth of the detected lines ranges from less than 0.1\% to about 50\%. To illustrate this diversity, we show a synthetic DIB absorption spectrum in Fig.~\ref{plot:synthetic_spectrum}, using a list of detected DIBs from \citet{Jenniskens1994} who studied high signal-to-noise ratio (S/N) spectra of four early-type stars. This synthetic spectrum illustrates that the DIB absorption is weak:  the strongest feature at $\lambda=4430\,$\AA\ has an equivalent width of only about $0.2\,$\AA\  for a dust column density corresponding to  \ebv$\simeq0.1\,$mag. 

For several decades, the main probe of DIB absorption has been high-S/N spectra of hot stars, in which weak absorption features can be measured and blending with stellar lines is minimized.
Dedicated surveys have included a few thousand stars at most \citep[e.g.,][]{vanLoon2013}
or only around a hundred stars if weaker DIBs are targeted \citep[e.g.,][]{Friedman2011}. In recent years, the availability of generic large sky surveys has allowed the study of DIBs in a statistical context. In addition to increasing the number of available lines of sight, they have also motivated the use of other strategies to detect DIBs in spectra of a wider range in spectral types.
Using the Sloan Digital Sky Survey \citep[SDSS;][]{YorkSDSS}, \citet{Yuan2012} reported the detection of two DIBs, $\lambda$5780 \& $\lambda$6283, in about $2,000$ stellar spectra (of all types) and characterized their strengths and radial velocities. Using the Radial Velocity Experiment \citep[RAVE;][]{SteinmetzRave}, \citet{Kos2013,Kos2014} detected DIB $\lambda8620$ below the noise level of individual spectra. They characterized its absorption in the composite spectra of several thousand cool stars and mapped the absorption in the sky. Using near-infrared spectra from the SDSS Apache Point Observatory Galactic Evolution Experiment \citep[APOGEE;][]{Majewski2014}, \citet{Zasowski2014} studied the DIB absorption at $\lambda=1.527\,\micron$ in about 100,000 stellar spectra probing a wide range of Galactic environments and mapped out its properties.

%----------------------------------
% figure
%----------------------------------
%----------------------------------

In this paper
\footnote{While performing this analysis we became aware of a similar effort by \citet{Baron2014}. We decided to finish the two analyses independently and submit the two papers to arxiv.org simultaneously.}
we use all types of spectra taken by the SDSS I, II, and III, i.e. lines of sight towards stars, galaxies and quasars, to map out the distribution of DIBs induced by the Milky Way. We show that, after carefully removing the intrinsic spectral energy distribution (SED) of each source, it is possible to measure statistical flux fluctuations at the $10^{-3}$ level. We then detect and characterize a set of 20 DIBs and study their correlations with various ISM tracers. Throughout the paper, we use air wavelength.

%% =======================================
\section{Data analysis}
%% =======================================
Our analysis makes use of optical spectra from the SDSS I, II \& III surveys with spectral resolution about 2000. We explore the detectability of DIB absorption in three types of sources: quasars, galaxies and stars. In each case we create absorption spectra, normalizing the observed source spectra by an estimate of the SED intrinsic to the source. In addition, we take extra care to handle artificial residuals originating from imperfections in the calibration process of SDSS spectra. For each type of sources, we use a different strategy to estimate the corresponding absorption spectra. To avoid contamination from atmospheric emission lines, mostly due to OH and ${\rm H_2O}$, we restrict our analysis to wavelengths below $6700 \rm \, \AA$. We now describe the analysis procedure applied to each type of sources. The next section will then present the characterization of absorption lines detected in the absorption spectra.

%----------------------------------
% figure
%----------------------------------
\begin{figure*}[h]
\begin{center}
\includegraphics[scale=0.4]{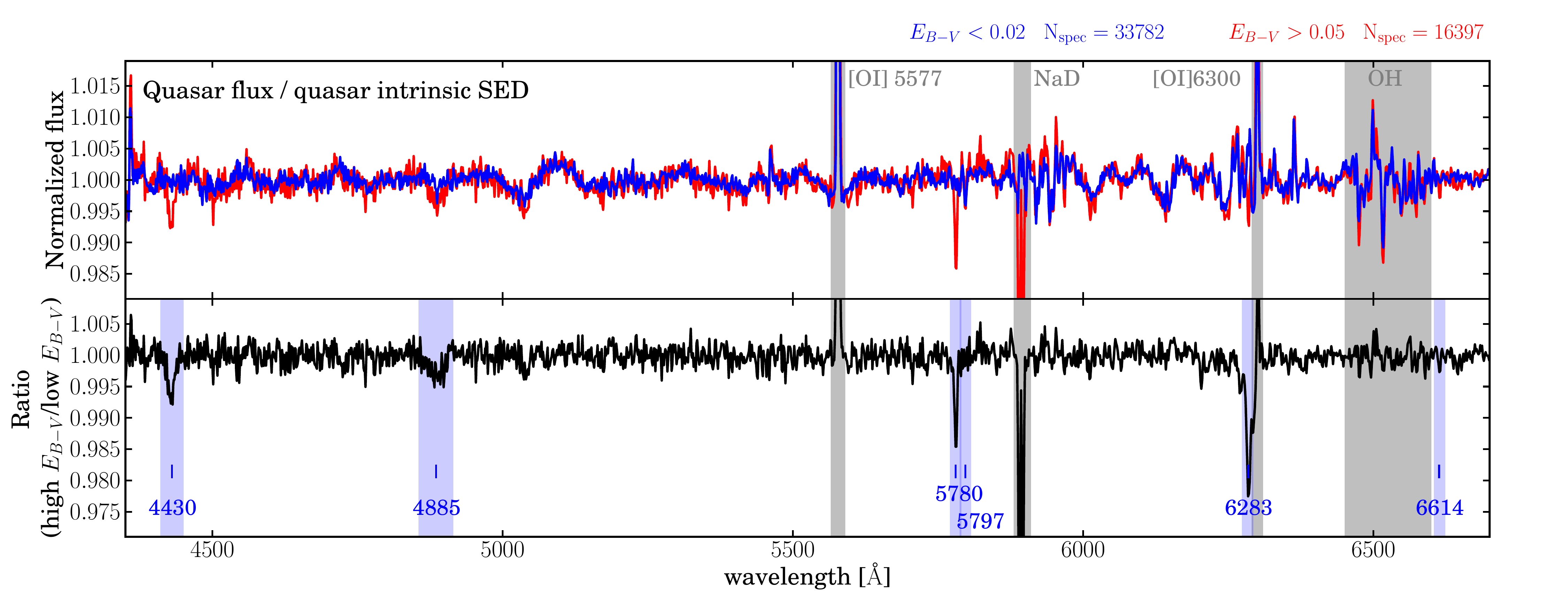}
\includegraphics[scale=0.4]{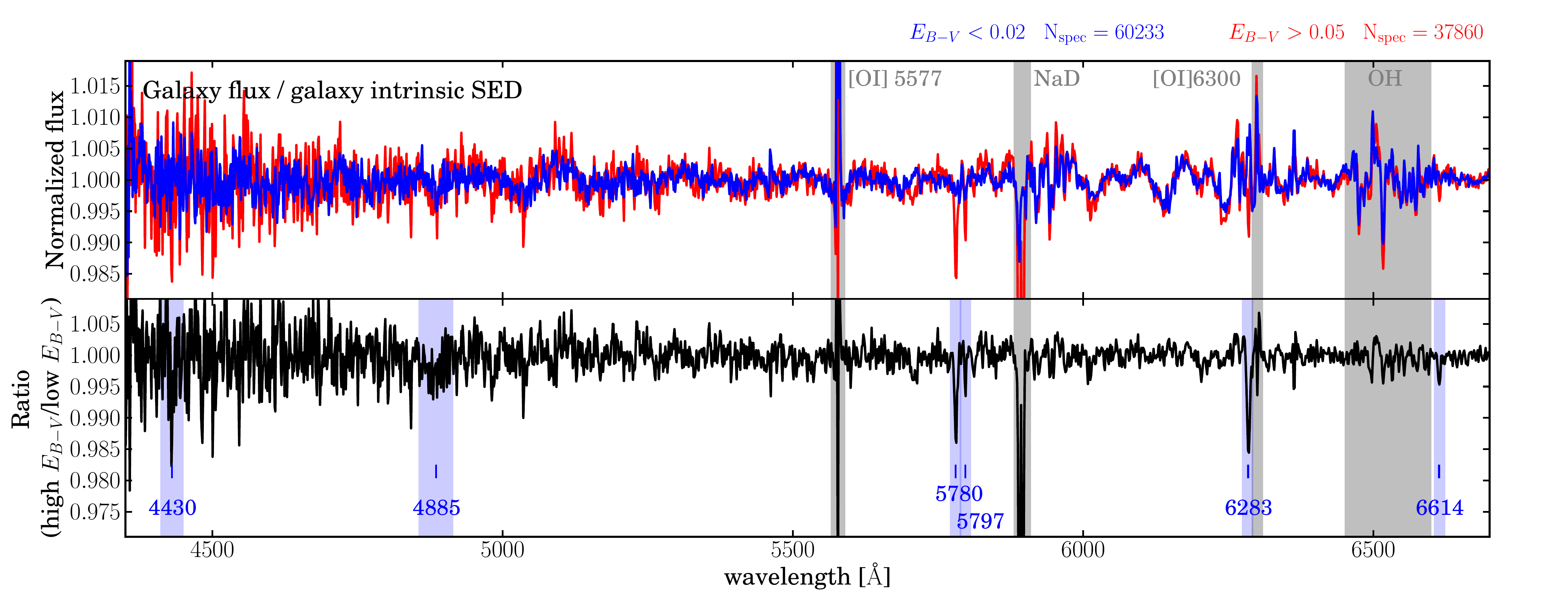}
\includegraphics[scale=0.4]{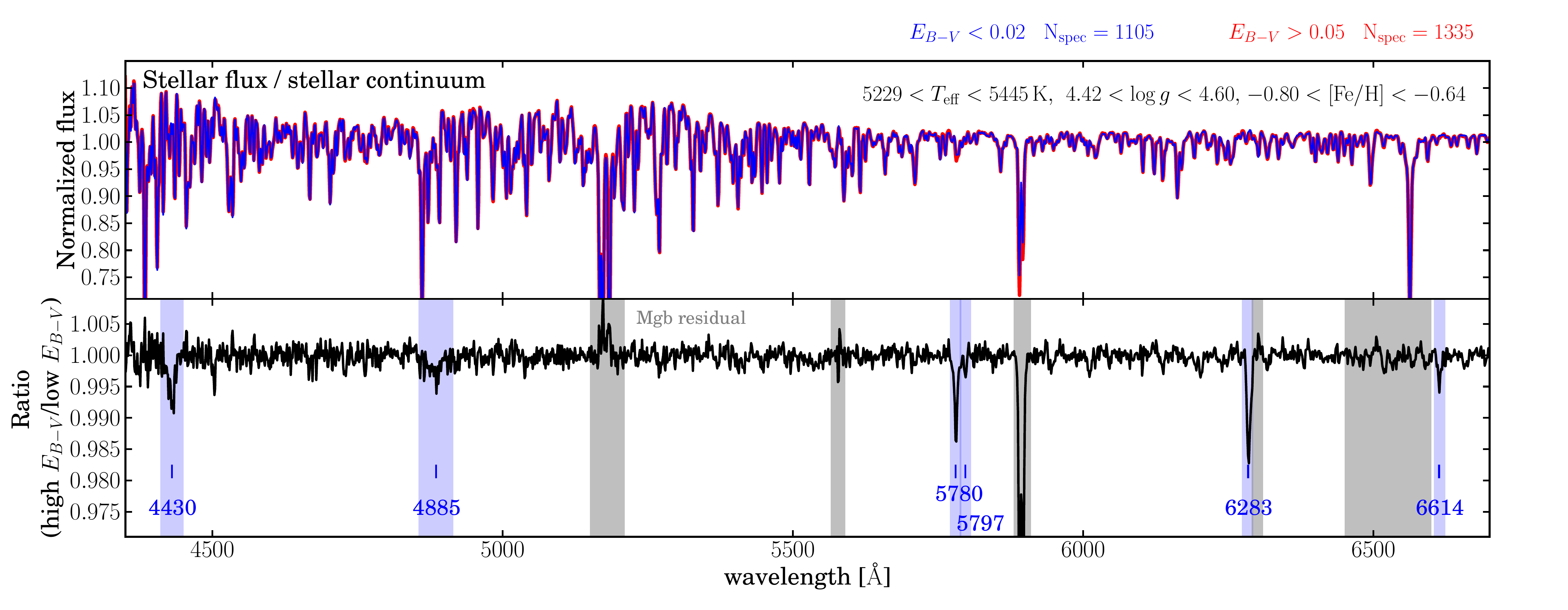}
\caption{
Illustration of the steps involved in the creation of composite absorption spectra using quasars (top), galaxies (middle) and stars (bottom). For each type of source the upper part of the panels shows continuum-normalized composite spectra for high and low dust reddening values and the bottom part of the panels shows their corresponding ratios. The intrinsic SED of quasars and galaxies are estimated through rest-frame template fitting. The continuum-normalized spectra show fluctuations imprinted by imperfections in the SDSS spectroscopic calibration. These features are detailed and explained in Appendix~\ref{appendix_a}. The intrinsic SED of stars is estimated using a running median filter to characterize the large-scale continuum. Intrinsic absorption features are not modelled. Their contribution between low and high dust reddening regions is expected to be comparable and has therefore close to no effect in the final ratio estimate. Blue bands indicate known DIBs. Gray bands indicate artefacts due to the presence of strong sky emission lines.}
\label{plot:methods}
\end{center}
\end{figure*}
%----------------------------------

\subsection{Creating absorption spectra}

\subsubsection{Quasars}
\label{sec:quasar}

We use the SDSS DR7 quasar catalogue compiled by \citet{SchneiderQSO}. The corresponding spectra were analysed by \citet{Zhu2013} who estimated their intrinsic continuum level. They did this using a dimensionality reduction technique (non-negative matrix factorization) to obtain a basis set of `eigen spectra' which can then be used to estimate each quasar's SED. We create absorption spectra by dividing each observed spectrum by its estimated intrinsic SED. We study corresponding spectra in the observer frame. 

We first create median composite absorption spectra for quasars observed in different regimes of Galactic dust reddening, derived from \citet[][hereafter SFD]{SFD1998}. In the upper part of the top panel of Fig.~\ref{plot:methods}, we show the results obtained from objects selected with \ebv$<0.02$ mag in blue and \ebv$>0.05$ mag in red. As can be seen, the composite spectra reveal systematic fluctuations at the $5\times10^{-3}$ level. These fluctuations appear to be inconsistent with Poisson noise and their amplitude does not correlate with dust column density. These features are mainly due to systematics in the spectral calibration process in SDSS as well as sky residuals. We discuss these effects in more detail in Appendix~\ref{appendix_a} and identify the origin of each of these features.

In order to overcome the limits given by the precision of the SDSS spectral calibration process we use the fact that DIB absorption is known to correlate with dust reddening (e.g., \citealt{Friedman2011, Welty2013}, and Section~\ref{sec:reddening}), while spectral calibration problems are not expected to do so. We can then improve the sensitivity of our flux residuals by considering the ratio between absorption spectra at high and low dust reddening regimes. To do so we consider the composite spectrum of quasars selected from low-reddening regions with \ebv$<0.02$ mag, where DIB absorption is expected to be small, as a reference spectrum. The ratio between the composite spectrum with high Galactic reddening and the reference composite spectrum with low Galactic reddening is shown in the lower part of the top panel of Fig.~\ref{plot:methods}. As expected this process removes features associated with the SDSS reduction process and provides us with an absorption spectrum with a scatter of the order of $1\times10^{-3}$. 
In the figure, the blue bands indicate the locations of several known DIBs.
The number of quasar spectra used in our analysis is listed in Table~\ref{table:number_spec}.

\subsubsection{Galaxies}

We select a set of luminous red galaxies from SDSS DR7. Such galaxies have well-defined SEDs. We use estimates of their intrinsic continuum fluxes provided by \citet{Zhu2010}. These authors modelled the observed galaxy spectra using single stellar population (SSP) models of \citet{BC2003} with the Padova 1994 library of stellar evolution tracks\footnote{\url{http://pleiadi.pd.astro.it/}}\citep[e.g][]{Girardi2000} and the \citet{Chabrier2003} initial mass function (IMF). We select galaxies at redshift greater than $0.2$. The number of galaxy spectra used in this analysis is listed in Table~\ref{table:number_spec}.

Similarly to the procedure used for quasars,  we create median composites of absorption spectra for different ranges of Galactic dust reddening. As shown in the upper part of the middle panel of Fig.~\ref{plot:methods} we find fluctuation patterns due to the spectroscopic calibration and sky emission/absorption features consistent with those obtained with quasar spectra. As done above, we use ratios of composite spectra to overcome these limitations, considering lines of sight with \ebv$<0.02$ mag as reference ones. The final ratio spectrum is shown in the lower part of the panel and allows us to detect a comparable set of DIBs.

%-------------- table --------------------------------
\begin{table}[t] 
\centering
\caption{Number of spectra}
\begin{tabular}{cccc}
\hline
Source & total  & reference sightlines & targeted sightlines \\
type & number & $E_{B-V}<0.02$ mag & $E_{B-V}>0.02$ mag\\
\hline
Quasar & 105,782 & 33,782 & 72,000\\
Galaxy & 210,726 & 60,233 & 150,493\\
Star & 354,231 & 84,406 & 269,825\\
\hline
\end{tabular} 
\label{table:number_spec}
\end{table} 
%---------------------------------------------------

\subsubsection{Stars}
\begin{figure}
\begin{center}
\includegraphics[scale=0.2]{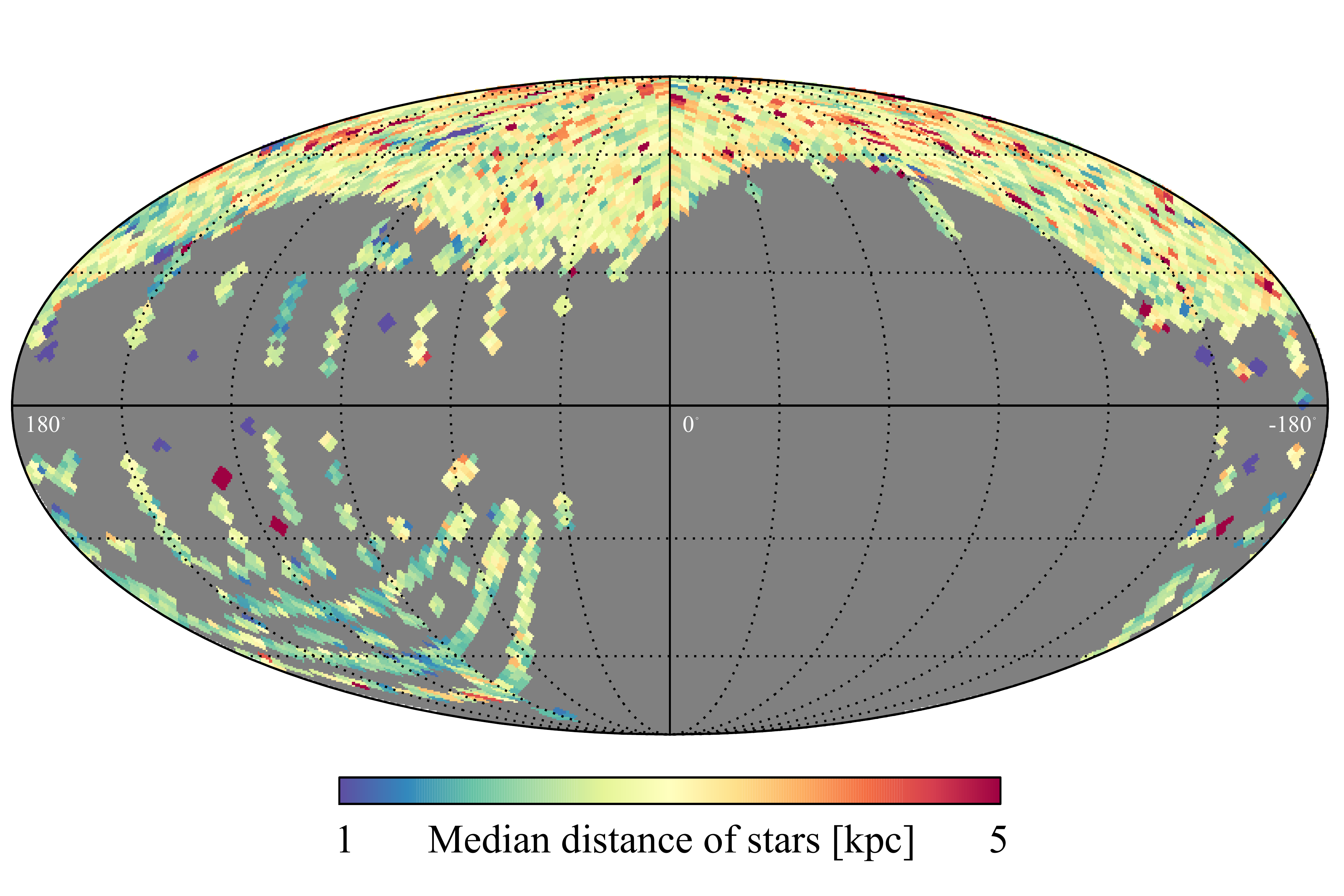}
\caption{Median distance of stars in the SDSS sky coverage. The typical distance is about 2--3 kpc, depending mildly on Galactic latitude.}
\label{plot:stellar_distance}
\end{center}
\end{figure}

We use the stellar spectra collected by the SDSS SEGUE I $\&$ II surveys \citep{Yanny2009,SDSSDR8}. We estimate the intrinsic SED of each star using a data-driven approach. We first remove large-scale fluctuations due to blackbody emission as well as the effect of line-of-sight dust extinction using a running median filter of size 200 pixels. To create absorption spectra and normalize out the contribution due to the intrinsic SED of each source, for a given star, we search for a set of reference stars at low dust reddening with \ebv$<0.02$ mag with similar stellar parameters. To do so we make use of three parameters: effective temperature ($T_{\rm eff}$), surface gravity ($\log g$) and metallicity ($\rm [Fe/H]$), estimated by the SEGUE SSPP pipeline \citep{Lee2008,Smolinski2011}
\footnote{\url{https://www.sdss3.org/dr10/spectro/sspp.php}}.
We construct a 3-D grid spanning the full range of the corresponding values: $4200<T_{\rm eff}<8700$\,K, $0.7<\log g<4.7$, and $-4.3< \rm [Fe/H]< 0.6$, with a resolution set to be four times lower than the dispersion of the stellar parameters for reference stars in each dimension. The corresponding bin sizes are $\rm 216\,K$ in $T_{\rm eff}$, 0.18 dex in $\log g$, and 0.16 dex in $\rm [Fe/H]$. These bin sizes are about the size of the systematic error of these three parameters. The SSPP pipeline estimated the stellar parameters based on several methods. We use the parameters derived from the ANNRR method \citep{ANNRR2007} which is based on continuum-normalized spectra. We also perform our analysis by using the parameters derived from other methods and find consistent results. In addition, the SSPP pipeline also provides distances of stars. In Fig.~\ref{plot:stellar_distance}, we show the median distances of stars within each sky pixel. The typical distance of stars is about 2--3 kpc.

We match each star to a set of reference stars from the same cell in the grid of stellar parameters. To reduce the effect of outliers or problematic spectra, we discard stars located in the stellar parameter bins with less than 20 corresponding stars with \ebv$<0.02$ mag. We also remove stars from SDSS plates with bad qualities
\footnote{\url{http://www.sdss3.org/dr8/algorithms/segueii/plate_table.php}}.
We then create a median composite spectrum for the reference stars. As an illustration, the upper part of the bottom panel of Fig.~\ref{plot:methods} shows the composite spectra of typical stars observed in SDSS at high \ebv\ (red) and low \ebv\ (blue) with the same stellar parameters.

To create an absorption spectrum we take the ratio between a stellar spectrum and its corresponding reference composite spectrum in the stellar rest frame. The lower part of the panel shows the ratio spectrum between the high \ebv\ composite spectrum and low \ebv\ reference composite spectrum. By applying our method, we are able to remove stellar absorption features effectively and detect strong DIBs, as indicated by the blue vertical bands. We note that this stellar residual spectrum makes use of only $0.5\%$ of all the available stellar spectra, while we have used all quasar and galaxy spectra (with \ebv$>0.05$ mag) in the top panels. This shows that most of the statistical power to map out the distribution of DIBs lies in the SDSS stellar spectra. Below we will therefore derive most of our results from stellar spectra. We will use quasar and galaxy spectra primarily for consistency checks. The total number of stellar spectra used in this analysis is listed in Table~\ref{table:number_spec}. More than $95\%$ of the stars used are F, G, and K stars with $4500<T_{\rm eff}<7000 \rm \, K$.

%----------------------------------
% figure
%----------------------------------
%----------------------------------

%----------------------------------
% figure
%----------------------------------
\begin{figure*}
\begin{center}
\hspace{-1cm}
\includegraphics[scale=0.38]{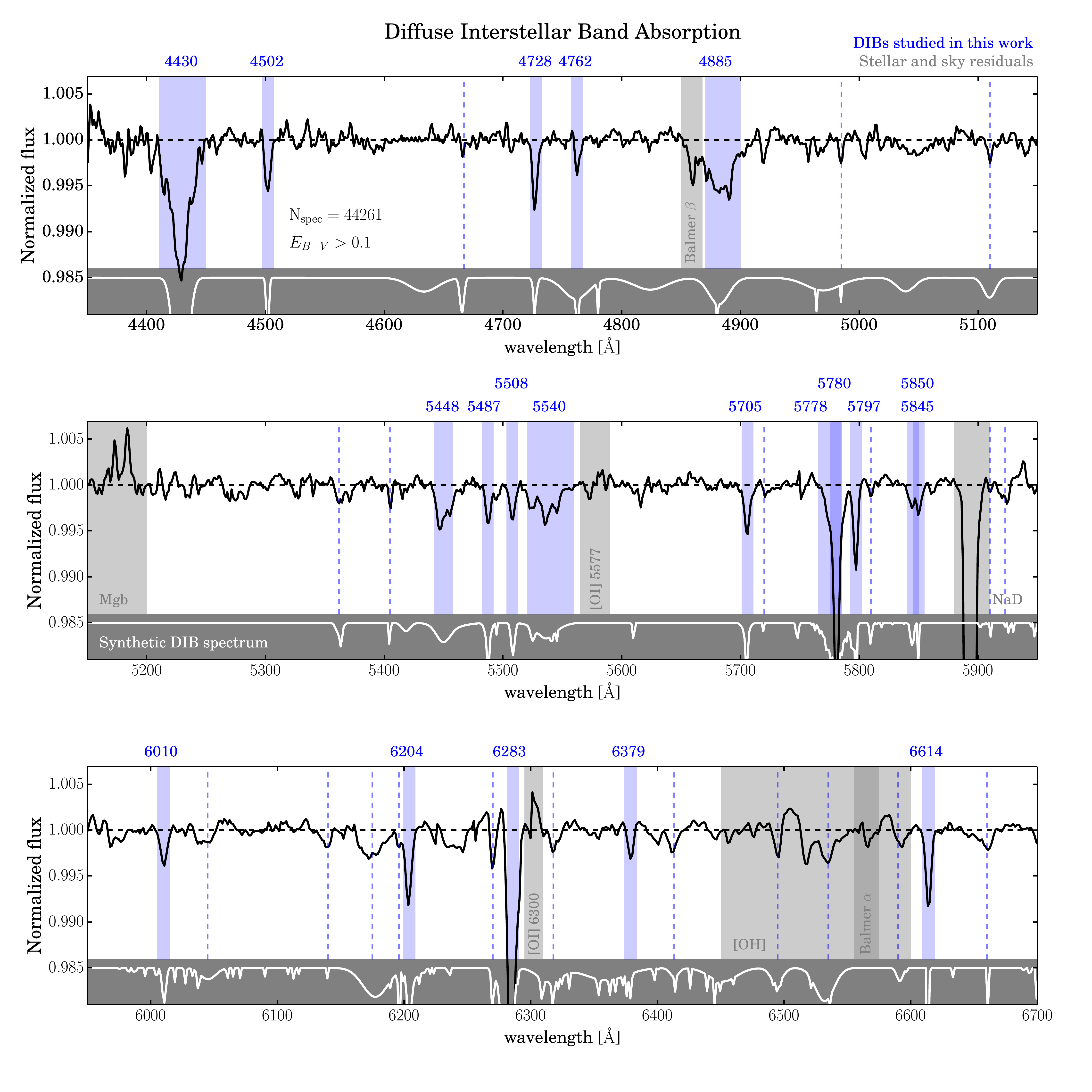}
\caption{An example of final composite absorption spectrum, combining more than $40,000$ stellar spectra at \ebv$>0.1$ mag. The light blue vertical bands indicate the 20 DIBs studied in this work, detected with more than $5\sigma$ and for which the line profile can be well characterized. The vertical dashed lines show weaker and/or broader DIBs detected but not used in the statistical analysis. The grey bands show residuals from sky lines and stellar absorption features. The white spectrum at the bottom shows a synthetic DIB absorption spectrum, as shown in Fig.~\ref{plot:synthetic_spectrum}.}
\label{plot:detection_example}
\end{center}
\end{figure*}
%----------------------------------

%----------------------------------
% figure
%----------------------------------
\begin{figure*}[!ht]
\begin{center}
\includegraphics[scale=0.4]{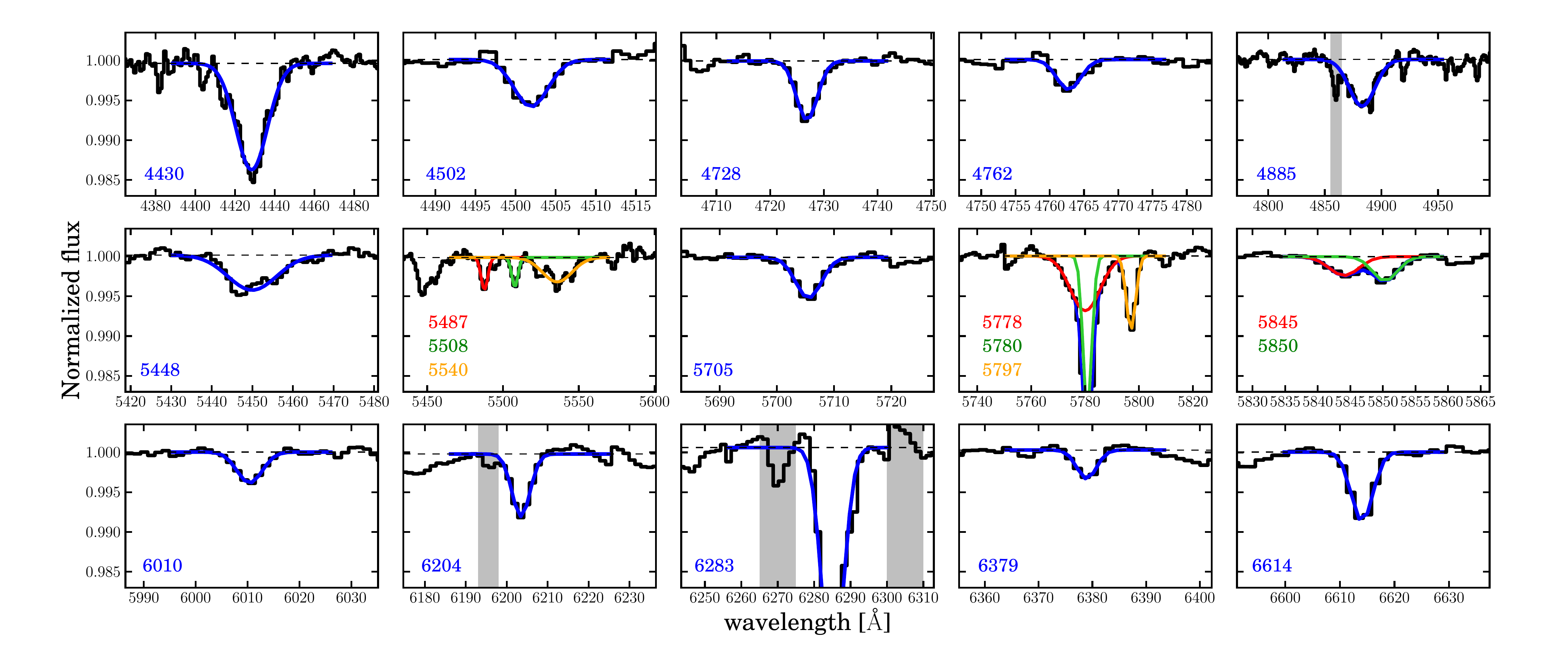}
\vspace{-0.6cm}
\caption{The selected set of 20 DIBs measured in a composite spectrum of 40,000 stellar spectra with \ebv$>0.1$ mag, as shown in Fig.~\ref{plot:detection_example}. Gaussian fits, used to estimate line parameters, are indicated with blue lines. For wavelength regions with multiple DIBs, we fit them simultaneously with a multiple-Gaussian profile, as shown with red, green, and orange lines. The grey bands indicate regions with contamination from nearby DIBs or sky/stellar residuals, which are masked in the fitting process.}
\label{plot:fitting}
\end{center}
\end{figure*}
%------------------

%----------------------------------
% figure
%----------------------------------
\begin{figure*}[ht]
\begin{center}
\includegraphics[scale=0.35]{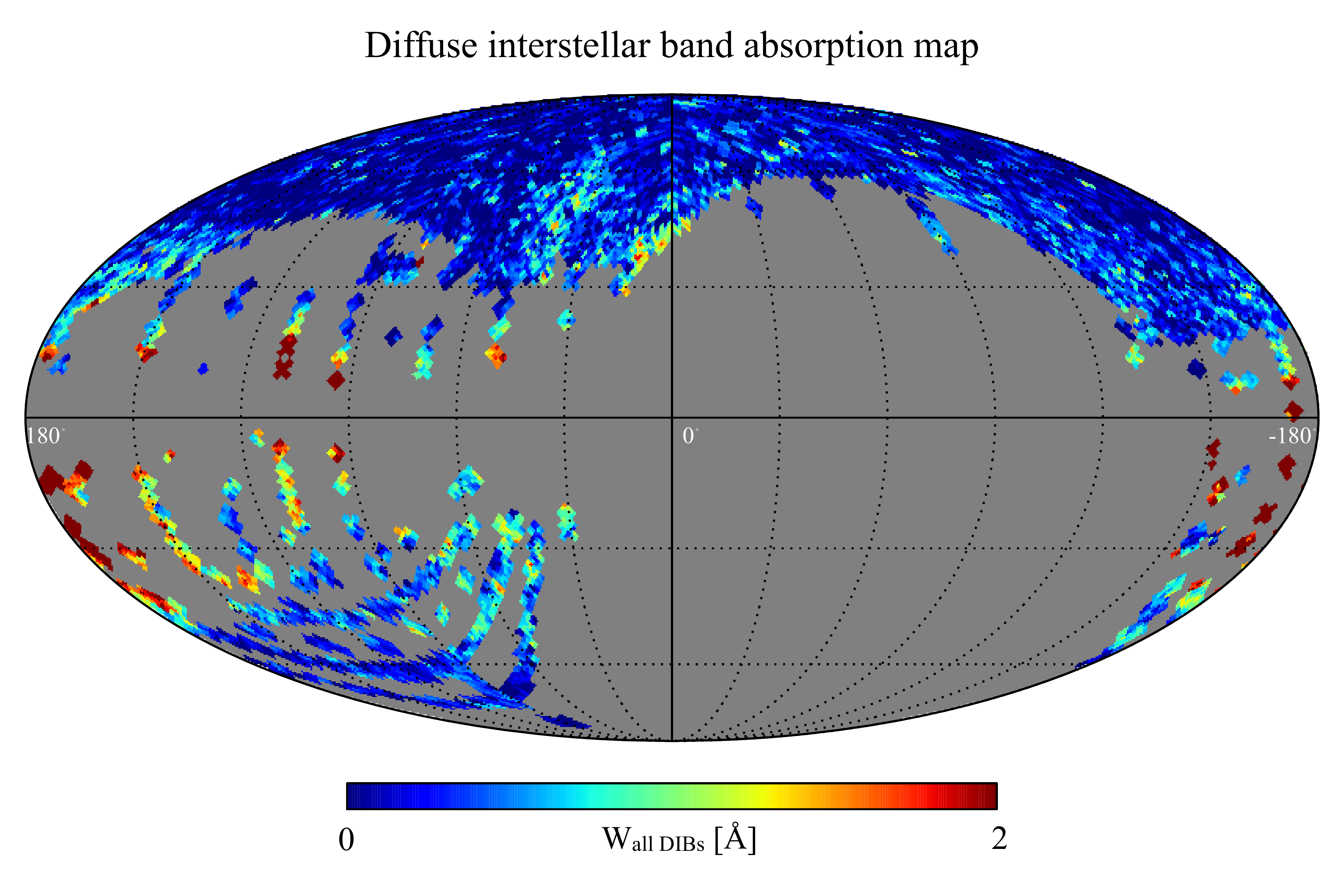}
\includegraphics[scale=0.35]{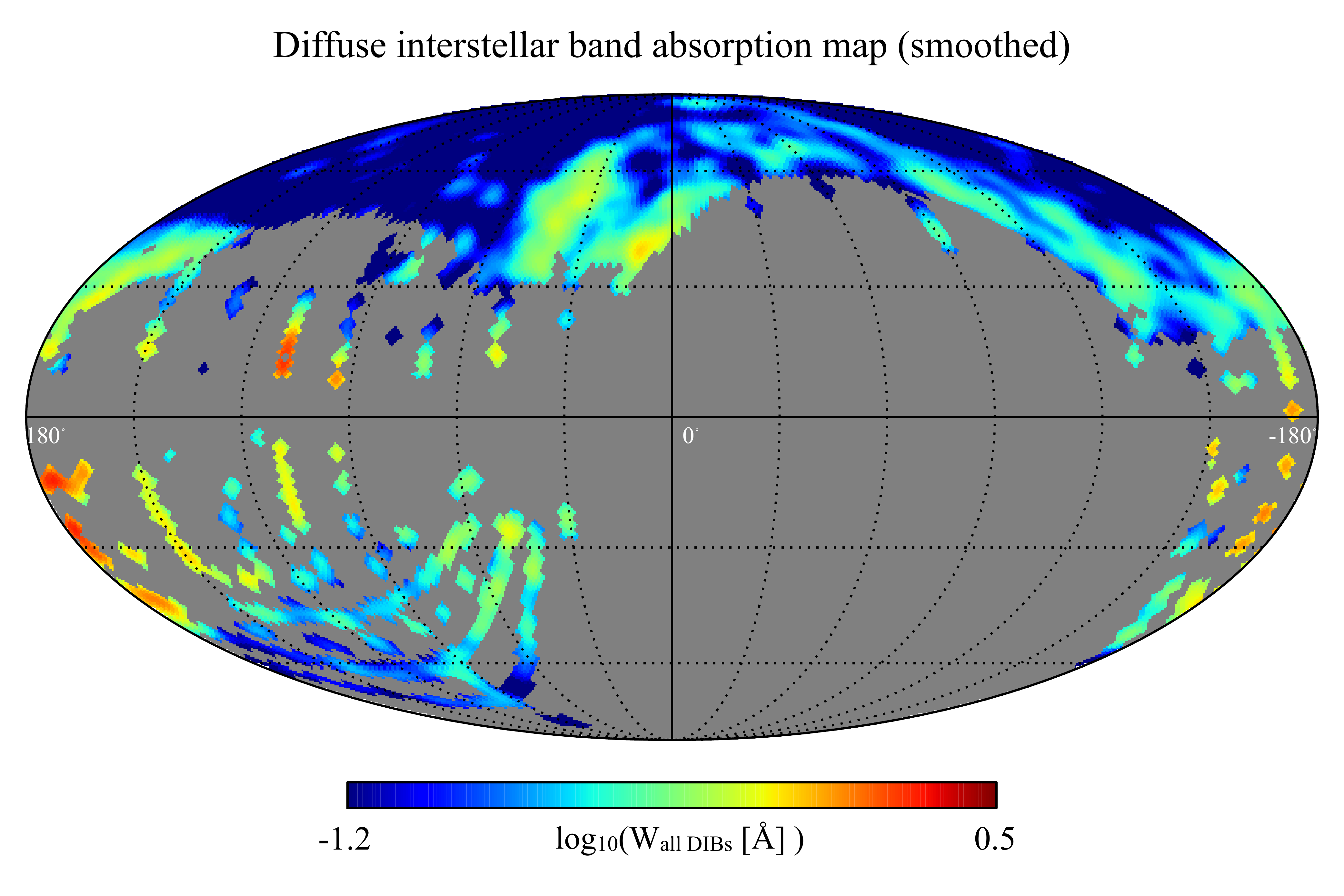}
\caption{\emph{Top:} full sky map of DIB absorption in Galactic coordinates estimated from more than 250,000 stellar spectra. Our correlation analyses are based on about 5,000 deg$^2$ corresponding to regions with $\ebv>0.02$ mag. Regions with lower dust column densities, i.e. $\ebv<0.02$ mag,  are used to define reference stars. The angular resolution of this map is about 1 deg$^2$ and the typical distance from stars to the Sun is about 2-3 kpc, mildly depending on latitude. Due to variation in the density of available spectroscopic data, some of the pixels of this map are signal or noise dominated.
\emph{Bottom:} smoothed version of the full sky map of DIB absorption, which is created by using a Gaussian kernel with FWHM$=5^{\circ}$. This map is signal dominated and reveals the morphology of DIB absorption in the sky. The map can be viewed interactively at 
\url{http://www.pha.jhu.edu/~tlan/DIB_SDSS/}.}
%%%% TWL: simplify the URLs
%dibs-map-Lan_et_al.html}.}}
\label{plot:DIB_map}
\end{center}
\end{figure*}
%----------------------------------

\subsection{Detection and characterization of DIBs}
\label{sec:DIB_define}

We now proceed to characterize DIBs in the composite residual spectra. Detecting and characterizing absorption lines require an accurate estimate of the continuum level. To do so we use the DIB catalogue compiled by \citet{Jenniskens1994} and create mask aimed at selecting the non-absorbed pixels of our spectra. From their list, we select DIBs with equivalent widths larger than $\rm 0.05\,\AA$ at \ebv$=1\,$mag but exclude broad DIBs with ${\rm FWHM} >25\,$\AA\ . For those, the continuum estimation is more difficult as the width of the absorption feature is an appreciable fraction of the median filter used for continuum estimation. We then mask out the wavelength regions corresponding to these DIBs and apply a median filter with a radius of 25 pixels to eliminate small-scale fluctuations in the residual spectra not accounted for in the previous steps of the analysis. Finally, we 
define the continuum level around each DIB by selecting wavelength regions uncontaminated by calibration errors or sky emission/absorption features.
As an illustration, Fig.~\ref{plot:detection_example} shows a composite absorption spectrum for all stars with \ebv$>0.1$ mag. The median \ebv\ of these stars is about $0.15$ mag. This absorption spectrum shows a set of 20 DIBs detected at more than $\sim5\sigma$, and not affected by residuals due to the sky lines and/or stellar absorption features. These absorption features are indicated with blue vertical bands and listed in Table~\ref{table:best_fit_power_law}. We can observe additional weak and/or broad absorption features which are consistent with known DIBs. They are indicated by blue dashed lines. Studying these weaker and broader features, however, requires a more detailed analysis to properly estimate the continuum level and the effect of possible artefacts in the flux residuals. We will therefore limit the present analysis to the above set of 20 DIBs with robust detections and characterization. 
The composite spectra from quasars and galaxies allow us to robustly detect DIBs $\lambda4430$, $\lambda4885$, $\lambda5780$, $\lambda6283$, and $\lambda6614$. We note that the SDSS footprint for extragalactic sources covers mostly high Galactic latitudes corresponding to low levels of dust reddening \ebv$<0.5$ mag. These types of sources do not allow us to probe a wide range of ISM column densities. We therefore only use them for consistency check.
We measure the equivalent width of DIBs with a Gaussian fitting of the line profile. For cases involving blended features, we make use of Gaussian profiles with multiple components. For example, around $\lambda=5780\,$\AA, three DIBs are known to exist: $\lambda5778$ (broad), $\lambda5780$ (narrow), and $\lambda5797$ (narrow). We measure these three DIBs simultaneously with a triple-Gaussian profile. For some broad DIB features due to blending of several weak DIBs, we fit a single Gaussian and quote them as a single DIB (e.g., $\lambda5540$). To identify DIBs potentially blended with multiple components, we compare the 20 DIBs with the DIB list\footnote{\url{http://dibdata.org/HD204827/}} compiled by \citet{Hobbs2008} with 8km/s spectral resolution and search for DIBs identified as blended. In Table~\ref{table:best_fit_power_law}, we mark those blended DIBs with star symbols.
For those blended DIBs, the derived correlations in this analysis can be driven by either a single dominant DIB or a combination of multiple DIBs with different dependences. For example, DIB$\lambda5540$ in our analysis consists of three narrow DIBs with two of them, $\lambda5541$ and $5546$ correlated with $\rm C_{2}$ molecules \citep{Thorburn2003}. 

We show examples of such line-profile fits in Fig.~\ref{plot:fitting}. The black histogram shows the composite absorption spectrum as shown in Fig.~\ref{plot:detection_example} with \ebv$>0.1\,$mag and the blue lines show the best-fitting single-Gaussian profiles used to estimate the absorption equivalent width. Multi-component fits are shown with red, green, and orange lines.
When measuring the equivalent widths of DIBs from high-S/N composite spectra, we fit simultaneously for the width, centre, amplitude, and continuum of the absorption spectra. However, when considering lower-S/N composite spectra, which is the case for quasars, galaxies and the composite spectra of a small number of stars, we fix the width of the Gaussian profiles estimated from high-S/N composite spectra but allow other parameters to vary. From high-S/N composite spectra, we find the width of each DIB does not vary with the Galactic dust reddening or other variables such as sky position. Fixing the width therefore allows us to estimate the equivalent width robustly.

% % % % % % % % % % % % % % % % % % % % 
\section{Results}
% % % % % % % % % % % % % % % % % % % % 

%----------------------------------
\subsection{The DIB absorption map}
\label{sec:sky}
%----------------------------------

Having characterized the absorption of 20 DIBs, we can map out the strength of each band as a function of position in the sky and then use them for cross-correlation analyses with various tracers of the ISM. We first investigate the overall spatial distribution of DIB absorption. To do so we pixelize the sky under the HEALPix\footnote{\url{http://healpix.sourceforge.net/}}\citep[][]{Gorski2005} scheme in the Galactic coordinate system. The resolution of the map can be adjusted given the purpose of the analysis. We locate all the stars in each pixel, create a composite spectrum and measure the strength of each of the 20 selected DIBs following the procedure described in the previous section. We choose the number of  HEALPix pixels along the Galactic longitude to be $N_{\rm side}=64$, which divides the whole sphere to 49152 pixels with equal area of about 1 deg$^{2}$. This resolution is motivated by the surface number density of the observed stars such that, in the majority of the pixels, there are enough stars for the robust absorption measurements of individual DIBs. To ensure reasonable S/N for the characterization of DIBs we only consider pixels with more than five stars. In total, we use $5929$ pixels (out of $9516$), covering about $5000$ deg$^{2}$ of the sky. The typical S/N of the composite residual spectra (normalized to unity) is about $150$, and the typical error of a DIB equivalent width ranges from 20\,m\AA \ for narrow DIBs (e.g., DIB$\lambda4728$) to 40\,m\AA \ for broad DIBs (e.g., DIB$\lambda4430$).

We measure the absorption strength of each DIB in each pixel and create 20 such maps. 
To display the global DIB absorption on a map, we combine the total absorption signal from the 20 DIBs considered:
\begin{equation}
{\rm W_{all\ DIBs}} = \sum_{i=1}^{20} {\rm W_{DIB_{i}}}\,.
\end{equation}
In Fig.~\ref{plot:DIB_map}, we present the map of total DIB absorption. We note that the measurements in different pixels are quasi-independent from each other because some common reference stars are used to define the zero-points. The map shows that the DIBs are more concentrated towards the disc, even though the sampling is relatively sparse at low Galactic latitude. This is expected from their known correlation with dust. We also observe smaller-scale features corresponding to known structures and clouds in the Milky Way. For example,  at $l\sim 170^\circ$ and $b\sim-15^\circ$, we observe the Taurus molecular cloud (located at about $100\,$pc from the Sun) and its surroundings. While at $l\sim-160^\circ$ and $b\sim-20^\circ$, we are able to measure the distribution of the DIBs at the edge of the Orion molecular cloud.
In the next section, we cross-correlate these maps with other all-sky surveys and study the correlations of DIBs with various ISM tracers.

%----------------------------------
\subsection{Dependence on ISM tracers}
%----------------------------------
We now take advantage of the large-scale mapping of DIB absorption enabled by our analysis to investigate the dependence of DIBs on other ISM tracers. To do so we make use of four publicly available all-sky maps tracing metals and hydrogen:
\begin{itemize}

\item \textbf{Dust}: we use the map created by \citet[][SFD]{SFD1998}, based on $100\,\micron$ flux from the COBE/DIRBE and IRAS/ISSA maps. It provides estimates of dust column densities in units of \ebv\ reddening.
\item \textbf{Polycyclic aromatic hydrocarbons}: we use the emission map in the $12\,\micron$ channel of the WISE all-sky survey \citep{Wright2010}, provided by \citet{WISE12um2014}\footnote{\url{http://faun.rc.fas.harvard.edu/ameisner/wssa/}}. The WISE $12\,\micron$ channel traces PAHs giving rise to emission between $7$ and $18\,\micron$. We note that our analysis will only make use of derivatives of this map and its overall normalization is not relevant for our purposes.
\item \textbf{Neutral atomic hydrogen}: we use the Leiden/Argentine /Bonn (LAB) Galactic \hyi\ $21\,$cm Survey \citep{LABHI2005}\footnote{\url{http://lambda.gsfc.nasa.gov/product/irsa/fg_LAB_HI_Survey_get.cfm}}. The LAB Survey is one of the most sensitive Milky Way $21\,$cm survey to date, with the extensive coverage both spatially and kinematically. To compare with our integrated absorption along lines of sight, we use their total \Nhyi\ values integrated over the full velocity range from $-450\,\kms$ to $400\,\kms$. 
\item \textbf{Molecules}: we use CO as a proxy for molecules and use $\rm CO_{1-0}$ emission map provided by PLANCK \citep{PlanckCO2013}\footnote{\url{http://irsa.ipac.caltech.edu/data/Planck/release_1/all-sky-maps/}. We use the {\tt Type 2} map which has a better S/N than the {\tt Type 1} map.}. We convert the integrated line intensity $W_{\rm CO}$ to the column density of molecular hydrogen $N_{\rm H_2}$ using the CO-to-$\rm H_2$ conversion ``$X$'' factor suggested by \citet{BolattoCO2H2}:
\begin{equation}
X_{\rm CO} \equiv \frac{N_{\rm H_2}}{W_{\rm CO}} = 2\times10^{20}\, {\rm cm^{-2}}/({\rm K}\,\kms)\,.
\end{equation}
The statistical noise level of the CO map at $\sim15\,'$ resolution is about $0.45\,{\rm K}\,\kms$, corresponding to about $0.9\times10^{20}\, {\rm cm}^{-2}$.
\end{itemize}
We note that all the above maps are based on 
emission measurements which probe the ISM over substantial path lengths, including material in the Milky Way and nearby galaxies located behind the set of stars used in our absorption analysis. This effect is stronger at lower Galactic latitudes and limits our ability to measure \emph{absolute} relations between a given absorption band and ISM column density. 
The tendency to overestimate ISM column densities due to the background contamination results in increasing the scattering and slightly lowering the amplitude of the measured correlation compared to the intrinsic one.
However, it is important to note that this effect will affect all DIBs in the same way. Measuring changes in the correlations \emph{between different DIBs} and ISM tracers will reveal intrinsic differences in the environmental dependencies of the corresponding DIBs.

%----------------------------------
% figure
%----------------------------------
\begin{figure}[!t]
\begin{center}
\includegraphics[scale=0.48]{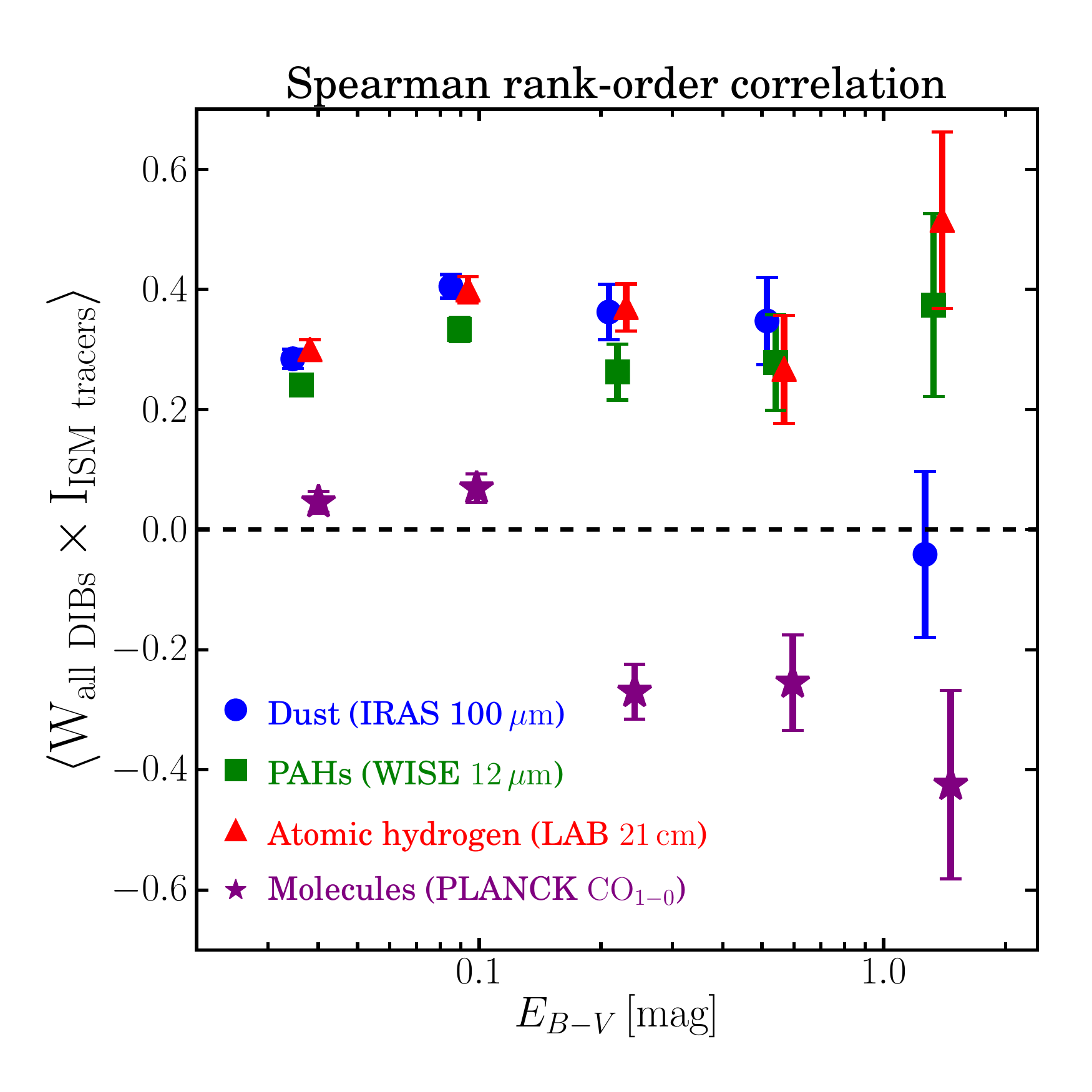}
\caption{Spearman rank-order correlation coefficients between the total DIB absorption and four ISM tracers as a function of \ebv. The dependence between DIBs and molecules appears to be different than that of other tracers.}
\label{plot:DIB_vs_ISM_correlation}
\end{center}
\end{figure}
%----------------------------------

%------------------------------------------------------
\begin{table}[ht] 
\caption{Number of spectra used in the equivalent width versus reddening relation}
\centering
\begin{tabular}{ccccc}
\hline\hline
$\langle E_{B-V} \rangle$ & $\rm N_{spec}$ & $\rm S/N$ \\ [0.5ex] % inserts table 
(mag) & &\\ [0.5ex]
%heading 
\hline
0.023 & 46052 & 3313 \\
0.031 & 52501 & 2527 \\
0.041 & 48285 & 2108 \\
0.054 & 40407 & 1647 \\
0.072 & 27651 & 1236 \\
0.097 & 18913 &  934 \\
0.127 & 15290 &  734 \\
0.170 & 7818 &  599 \\
0.230 & 4675 &  480 \\
0.305 & 3476 &  355 \\
0.408 & 2045 &  298 \\
0.548 & 1264 &  283 \\
0.728 &  710 &  268 \\
0.982 &  446 &  255 \\
1.247 &  211 &  239 \\
\hline
\end{tabular}
\label{table:high_SN_composite}
\vspace{.2cm}
\end{table} 
%----------------------------------
% figure
%----------------------------------
\hspace{-0.5cm}
\begin{figure*}[!ht]
\begin{center}
\includegraphics[scale=0.42]{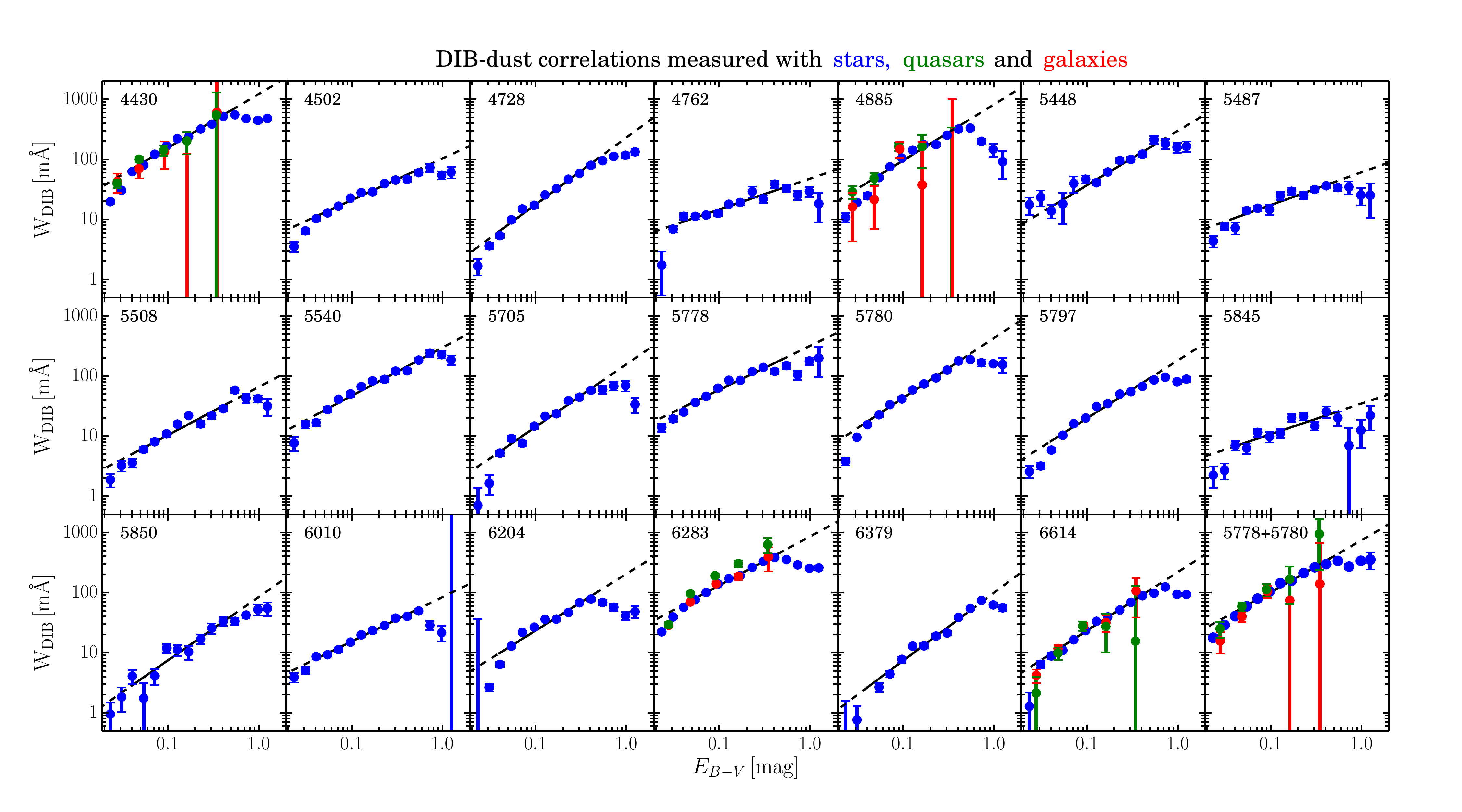}
\caption{Equivalent width measurements of 20 DIBs as a function of dust reddening in units of \ebv. Each equivalent width is measured from a high S/N composite spectrum combining thousands of spectra with similar Galactic reddening. Table \ref{table:high_SN_composite} shows the number of spectra used in the composite spectra and the S/N of the composite spectra. The blue, green, and red data points show median values measured in star, quasar, and galaxy spectra, respectively. The solid black lines are best-fitting power laws derived from measurements of stellar composite spectra at $0.04<E_{B-V}<0.5$ mag. The black dashed lines are extrapolations to higher and lower reddening values. We note that quasar and galaxy composite spectra do not have enough S/N to decouple DIBs $\lambda5778$ and $\lambda5780$. We therefore measure the sum of these two components and show they are consistent with measurements from stars.}
\label{plot:DIB_vs_reddening}
\end{center}
\end{figure*}
%----------------------------------

We first investigate the total DIB absorption field and explore its global dependence on each tracer introduced above. To do so we resample all the maps to the same resolution: $N_{\rm side}=64$ which corresponds to about 1 deg$^{2}$. To approach the problem generically we measure the Spearman rank-order correlation coefficient between the total DIB absorption field and each tracer introduced above. We show the amplitude of the corresponding correlations, in bins of \ebv, in Fig.~\ref{plot:DIB_vs_ISM_correlation}. As can be seen, we find positive correlations between DIB strength and dust, PAHs and atomic hydrogen. In contrast, the correlation coefficient with molecules appears to be negative at high dust column densities. This indicates that molecules have a different relationship with the total DIB absorption field from other tracers. We note that at low extinction, most $N_{\rm H_2}$ values are below the noise level, so the amplitude of the correlation coefficients in this regime mostly reflects this fact rather than the intrinsic correlation, which is unmeasurable in our data. The interpretation of the respective amplitudes needs to account for the intrinsic correlations between each ISM tracer. This is investigated in more detail below.

%----------------------------------
% figure
%----------------------------------
\begin{figure*}[!t]
\begin{center}
\includegraphics[scale=0.5]{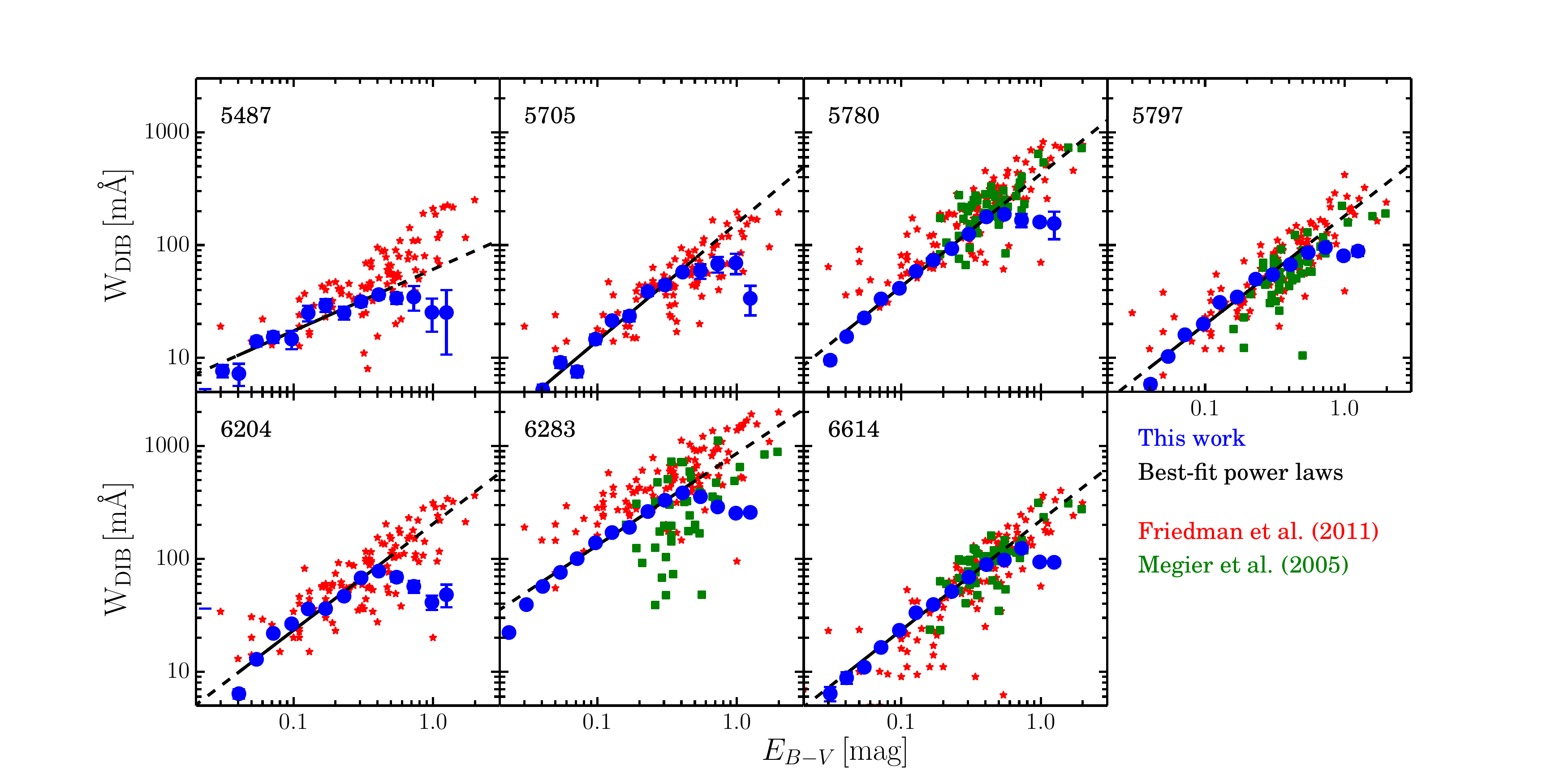}
\caption{Comparison between our statistical DIB absorption measurements (blue) and results obtained from studies of individual spectra of hot stars from \citet[][red]{Friedman2011} and \citet[][green]{Megier2005}. Note that the three sets of measurements involve different regions of the sky. The black line shows the best-fitting power-law trends estimated from our measurements with $0.04<E_{B-V}<0.5$ mag and extrapolated at the low and high ends.}
\label{plot:comparison}
\end{center}
\end{figure*}
%----------------------------------

%------------------------------------------------------
\begin{table}[ht] 
\caption{Best-fit parameters characterizing the
     relationships between DIBs and E(B-V) (equation~\ref{eq:w})}
\centering
\begin{tabular}{ccccc}
\hline\hline
$\lambda$ & A & $\gamma$ & ${\rm W/A_{V}}$ \\ [0.5ex] % inserts table 
(\AA) & & & (\AA/mag)\\ [0.5ex]
%heading 
\hline
$4430$ & $1.22 \pm 0.04$ & $0.89 \pm 0.02$  & 0.44 \\
$4502$ & $0.10 \pm 0.01$ & $0.69 \pm 0.03$  & 0.05 \\
$\ \, 4728^{*}$ & $0.23 \pm 0.01$ & $1.11 \pm 0.03$  & 0.07 \\
$4762$ & $0.05 \pm 0.01$ & $0.51 \pm 0.06$  & 0.03 \\
$\ \, 4885^{*}$ & $0.82 \pm 0.04$ & $0.93 \pm 0.02$  & 0.29 \\
$\ \, 5448^{*}$ & $0.30 \pm 0.04$ & $0.91 \pm 0.07$  & 0.11 \\
5487 & $0.06 \pm 0.01$ & $0.55 \pm 0.05$  & 0.03 \\
5508 & $0.07 \pm 0.01$ & $0.80 \pm 0.05$  & 0.03 \\
$\ \, 5540^{*}$ & $0.30 \pm 0.03$ & $0.80 \pm 0.04$  & 0.12 \\
$\ \, 5705^{*}$ & $0.16 \pm 0.01$ & $1.04 \pm 0.04$  & 0.05 \\
$\ \, 5778^{*}$ & $0.32 \pm 0.02$ & $0.73 \pm 0.03$  & 0.14 \\
5780 & $0.43 \pm 0.02$ & $1.00 \pm 0.02$  & 0.14 \\
5797 & $0.18 \pm 0.01$ & $0.96 \pm 0.02$  & 0.06 \\
$\ \, 5845^{*}$ & $0.03 \pm 0.01$ & $0.51 \pm 0.08$  & 0.02 \\
5850 & $0.08 \pm 0.02$ & $1.06 \pm 0.11$  & 0.03 \\
6010 & $0.08 \pm 0.01$ & $0.74 \pm 0.03$  & 0.04 \\
$\ \, 6204^{*}$ & $0.20 \pm 0.01$ & $0.94 \pm 0.02$  & 0.07 \\
$\ \, 6283^{*}$ & $0.86 \pm 0.02$ & $0.81 \pm 0.01$  & 0.34 \\
6379 & $0.10 \pm 0.01$ & $1.13 \pm 0.05$  & 0.03 \\
6614 & $0.22 \pm 0.01$ & $0.97 \pm 0.02$  & 0.07 \\
\hline
\end{tabular}
\begin{tablenotes}
\item *DIBs possibly blended with multiple weak DIBs
\end{tablenotes}
\label{table:best_fit_power_law}
\vspace{.2cm}
\end{table} 
%------------------------------------------------------

\subsubsection{Dependence on dust}
\label{sec:reddening}

It has long been known that the strength of DIBs correlates with the column density of dust \citep{Merrill1937}. Our statistical approach allows us to measure these dependencies using hundreds of thousands of lines of sight. To do so we select quasar, galaxy and stellar spectra as a function of dust column density from \citet{SFD1998}. This is done using DIB maps with a resolution comparable to that of the dust map. We combine spectra as a function of Galactic reddening to form high S/N median composite spectra and measure the equivalent widths of DIBs from the final composite spectra. 
The number of spectra used for the composite spectra and the corresponding S/N are shown in Table \ref{table:high_SN_composite}. The noise is the standard deviation of the spectra after removing outliers with 5-sigma clipping.
Fig.~\ref{plot:DIB_vs_reddening} shows the median equivalent width of the set of 20 selected DIBs in the previous section, as a function of dust reddening \ebv. The blue, green and red data points are measurements from stellar, quasar, and galaxy composite spectra. The error is estimated by bootstrapping each sample and represents the error of the median equivalent width.
Instead of constructing high S/N composite spectra, we also test the results by using the DIB equivalent width measured from each sky pixel with 1 deg$^{2}$ resolution and calculate the inverse variance-weighted mean DIB equivalent width as a function of \ebv. Two methods yield consistent results.

We find the equivalent widths of DIBs measured from the three different types of sources to be roughly consistent with each other. We note that a perfect agreement is not expected as the extragalactic sources have a different spatial distribution over the sky. As pointed out above, composite spectra from quasars and galaxies have lower S/N ratios. In some cases they do not allow us to decompose the blending of certain lines, for example between DIB $\lambda5778$ (broad) and $\lambda5780$ (narrow). In this case we therefore compare the sum of equivalent widths of the two DIBs measured from quasar and galaxy, stellar spectra (as shown in the lower-right panel). The consistency between the measured absorption strengths of DIBs from star, quasar, and galaxy spectra shows that the methods we applied effectively remove features intrinsic to the sources and the systematics in the spectral reduction. We note that the slightly higher equivalent width for DIB $\lambda6283$ estimated from composite quasar spectra is due to contamination from a nearby sky emission line. Such effect can be observed in Fig.~\ref{plot:methods}. 
%----------------------------------
% figure
%----------------------------------
\begin{figure}[h]
\begin{center}
\includegraphics[scale=0.48]{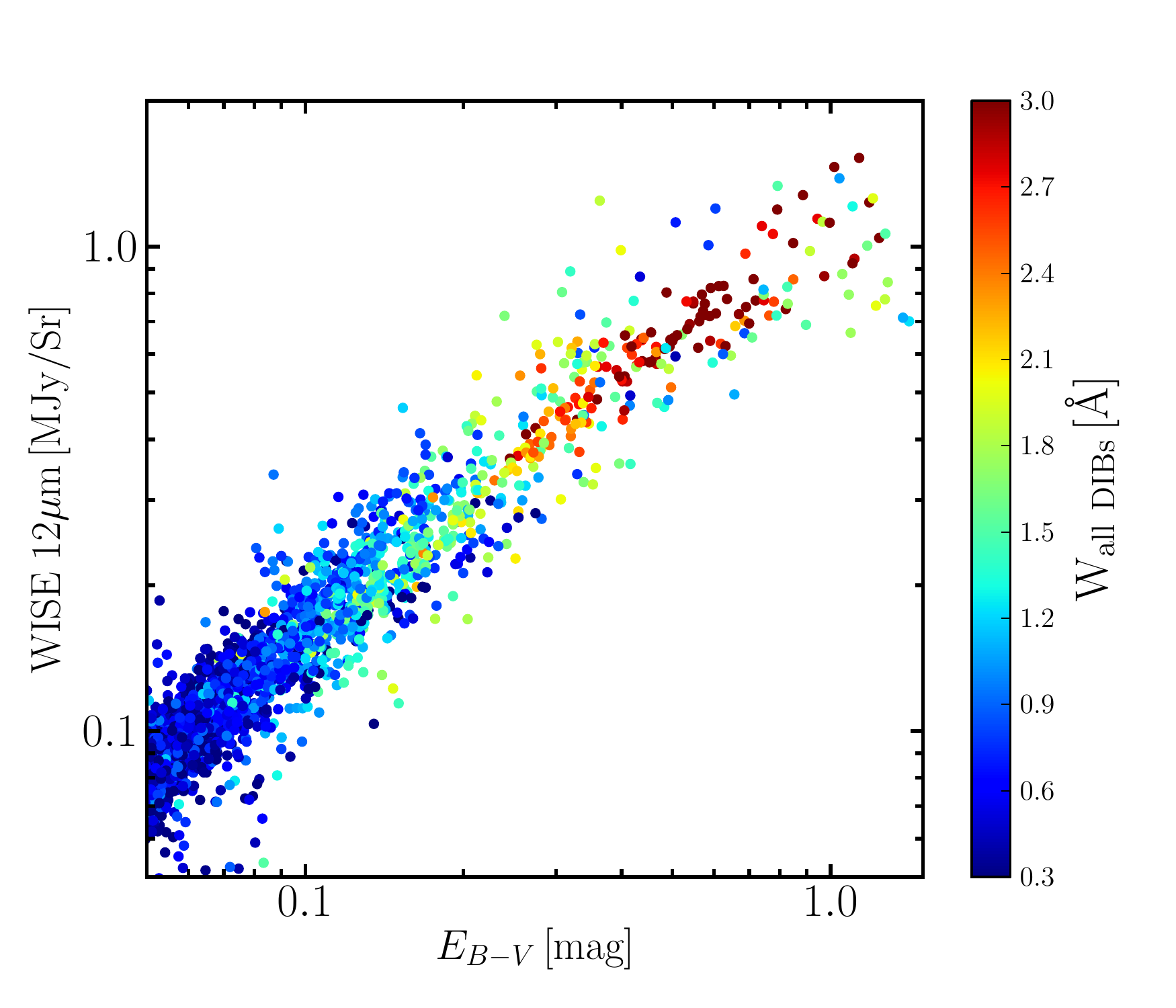}
\caption{Absorption strength of the 20 selected DIBs as a function of dust reddening and PAH emission measured by {\it WISE} in its $12\,\mu$m band. We find that PAHs and dust show a tight relation which does not allow us to disentangle their respective effects on DIB absorption.}
\label{plot:pah_ebv}
\end{center}
\end{figure}
%----------------------------------

Overall, our measurements confirm strong correlations between the strength of the 20 DIBs selected and Galactic reddening. In the regime \ebv$<0.5$, the median equivalent widths of all DIBs increase with reddening and the \emph{observed} relation between the two quantities can be well described by a power-law function form:
\begin{equation}
W_{\rm DIB} = A \times (E_{B-V})^{\gamma}\,. 
\label{eq:w}
\end{equation}

We fit our measurements with equation~(\ref{eq:w}) for stellar composite spectra with $0.04<$\ebv$<0.5\,$mag. The lower limit is selected to eliminate the effect of the observed departure from a power-law behaviour at low $\ebv$ values, which is due to two reasons: first we use lines of sight selected with $\ebv<0.02$ mag as reference lines of sight. This prevents us from measuring the absolute level of DIB absorption as we expect the measured value of $W_{\rm DIB}$ to be zero at $\ebv\lesssim0.02$ mag. Secondly, it is known that the SFD dust map (derived from infrared emission) is contaminated by the infrared emission of low-redshift galaxies \citep[see][]{Yahata2007}. As a result, $\ebv$ values lower than 0.04 mag ($A_{V}<0.1$ mag) tend to be overestimated. In the rest of our analysis we will focus primarily on the high reddening regime $\ebv>0.1$ mag and these effects can be safely neglected.

The black solid lines are the best-fitting power laws and the dashed lines are extrapolations towards high and low \ebv\ values. The best-fitting parameters are listed in Table~\ref{table:best_fit_power_law} and the relative strength of each DIB, $W_{\rm DIB}/A_V$, is estimated using a value of $R_V=3.1$ and
the best-fitting relations evaluated at $E_{B-V}=0.32$ mag (or $A_V=1$ mag). For high \ebv\ values, we observe different behaviours in the relation between absorption strength and dust reddening. The equivalent width of certain DIBs, for example DIBs $\lambda4502$, $\lambda4728$, $\lambda5540$, and $\lambda5850$ keeps increasing with dust column density. In contrast, other DIBs depart from the trends observed at lower column densities and flatten (e.g., $\lambda5780$) or even become weaker (e.g., $\lambda6283$). This behaviour, previously reported for a number of DIBs, has been referred to as the skin effect \citep{Herbig1995}. We emphasize that the different trends at high \ebv\ values reveal intrinsic differences in DIB behaviours. This relative effect is not affected by the fact that the SFD map provides estimates of dust column density across the entire Galaxy. The dust contribution associated with material located behind the set of stars is the same for \emph{all} DIBs. To demonstrate that, we also measure the Galactic reddening based on the difference between the observed g-r colour of a star and the colour of the reference and perform the same analysis. The detailed comparison is discussed in Appendix~\ref{appendix_b}. We found that although the SFD map tends to overestimate the Galactic reddening along the lines of sight, the correlations derived with $0.04<\ebv<0.5$ mag and the different behaviours at high Galactic reddening from two reddening estimators are consistent.

We now compare our measurements to other results from the literature. Fig.~\ref{plot:comparison} shows the dependencies for seven DIBs studied in \citet{Friedman2011} in red and four in \citet{Megier2005} in green. These studies were based on individual high-S/N spectra of hot stars. The blue data points correspond to the statistical measurements presented above. The black lines are our best-fitting power laws as shown in Fig.~\ref{plot:DIB_vs_reddening}. As can be seen, the different sets of measurements are overall in good agreement. While our measurements only show median values, the data points from \citet{Friedman2011} and \citet{Megier2005} show measurements for individual sightlines. We note that our estimated equivalent width for DIB $\lambda6283$ is slightly lower than that of \citet{Friedman2011}. This difference could be due to that the continuum estimate is affected by the nearby strong sky line. It is interesting to note that at the high end, certain DIBs display a different behaviour between the different analyses. We note that our sampling of the Galaxy differs from that of \citet{Friedman2011} and \citet{Megier2005}. The SDSS spectroscopic targeting generally avoided high-extinction regions. However, a number of special fields with known molecular clouds were specifically targeted towards high Galactic reddening regions. About $50\%$ of our lines of sight with $\ebv>0.5$ mag has molecular hydrogen fraction larger than 0.5 while the average molecular hydrogen fraction with $\ebv>0.5$ mag in the Milky Way is only about $15\%$.
Those lines of sight tend to intercept only one molecular cloud with high dust column density where the skin effect is mostly observed \citep{Herbig1995}.
These lines of sight are therefore not directly comparable to some of the environments probed in previous studies.

\subsubsection{Dependence on PAHs}

The global rank-order correlation coefficient shown in Fig.~\ref{plot:DIB_vs_ISM_correlation} indicates a positive correlation between the strength of the total DIB absorption field and the amount of PAHs traced by the WISE 12$\rm \,\micron$ band, similar to that found with the amount of dust. In Fig.~\ref{plot:pah_ebv} we show the relation between \ebv, $12\,\micron$ flux and the total DIB absorption $W_{\rm all \, DIBs}$, where $W_{\rm all\,DIBs}$ is indicated by the colour scale. First, we observe that the PAHs emission is directly proportional to that of the dust, with a scatter smaller than 0.3 dex. The strength of the total DIB absorption also appears to be roughly correlated with these two quantities. We cannot detect any vertical gradient in the DIB equivalent width at fixed \ebv\ value. In each \ebv\ bin the $W$ values appear to be symmetrically distributed around the mean value. This shows that the PAHs emission does not affect the observed values of DIB equivalent width beyond the effect already due to dust column density. In other words, our analysis does not allow us to disentangle the effects of dust and PAHs on the strength of the DIB absorption.

%----------------------------------
\subsubsection{Dependence on atomic and molecular hydrogen}
\label{sec:DIB_and_f_h2}
%----------------------------------

%----------------------------------
% figure
%----------------------------------
\begin{figure}[t]
\begin{center}
\includegraphics[scale=0.45]{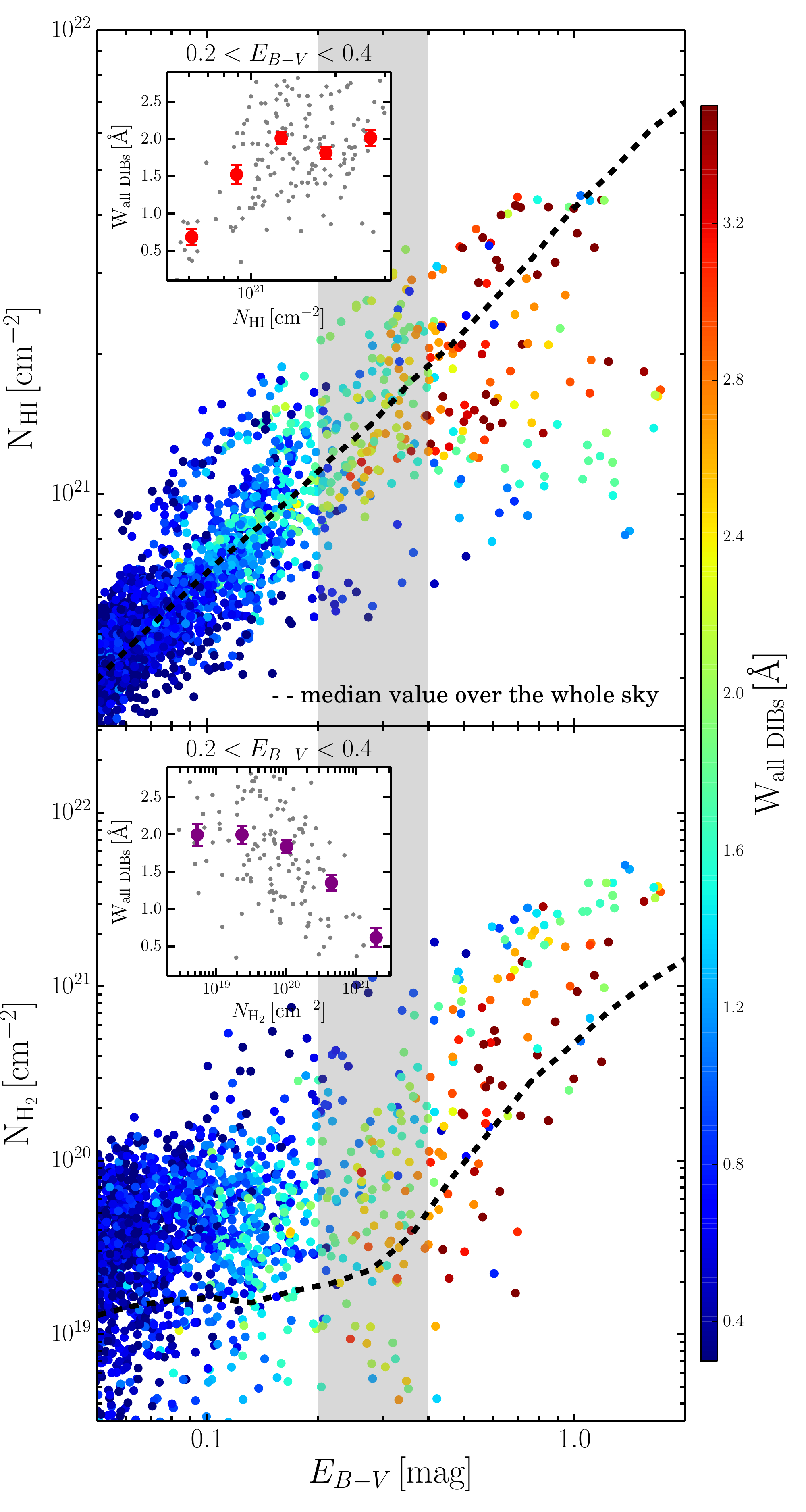}
\caption{Distribution of DIB absorption strength as a function of neutral hydrogen, molecular hydrogen, and dust column densities. \emph{Top:} projection on to the \Nhyi-\ebv\ plane. \emph{Bottom:} projection on to the \Nhytwo-\ebv\ plane. The insets show $W_{\rm all\,DIBs}$ as a function of \Nhyi\ (top) and \Nhytwo\ (bottom) in a narrow \ebv\ bin with $0.2<E_{B-V}<0.4$ mag shown in the grey regions. They illustrate the dependencies between parameters. The black dashed lines are the median of $\Nhytwo$ and $\Nhyi$ (including non-detection) as a function of $\ebv$ over the whole sky.  At $\ebv>0.5$ mag, most of our sightlines have $\Nhytwo$ higher than the median value over the sky because of the SDSS selection. Those lines of sight tend to have lower $\Nhyi$ than the median over the sky.The data points with error bars are the mean with standard errors.}
\label{plot:DIB_vs_H_H2}
\end{center}
\end{figure}
%----------------------------------

%----------------------------------
% figure
%----------------------------------
\begin{figure*}[ht]
\begin{center}
\includegraphics[scale=0.45]{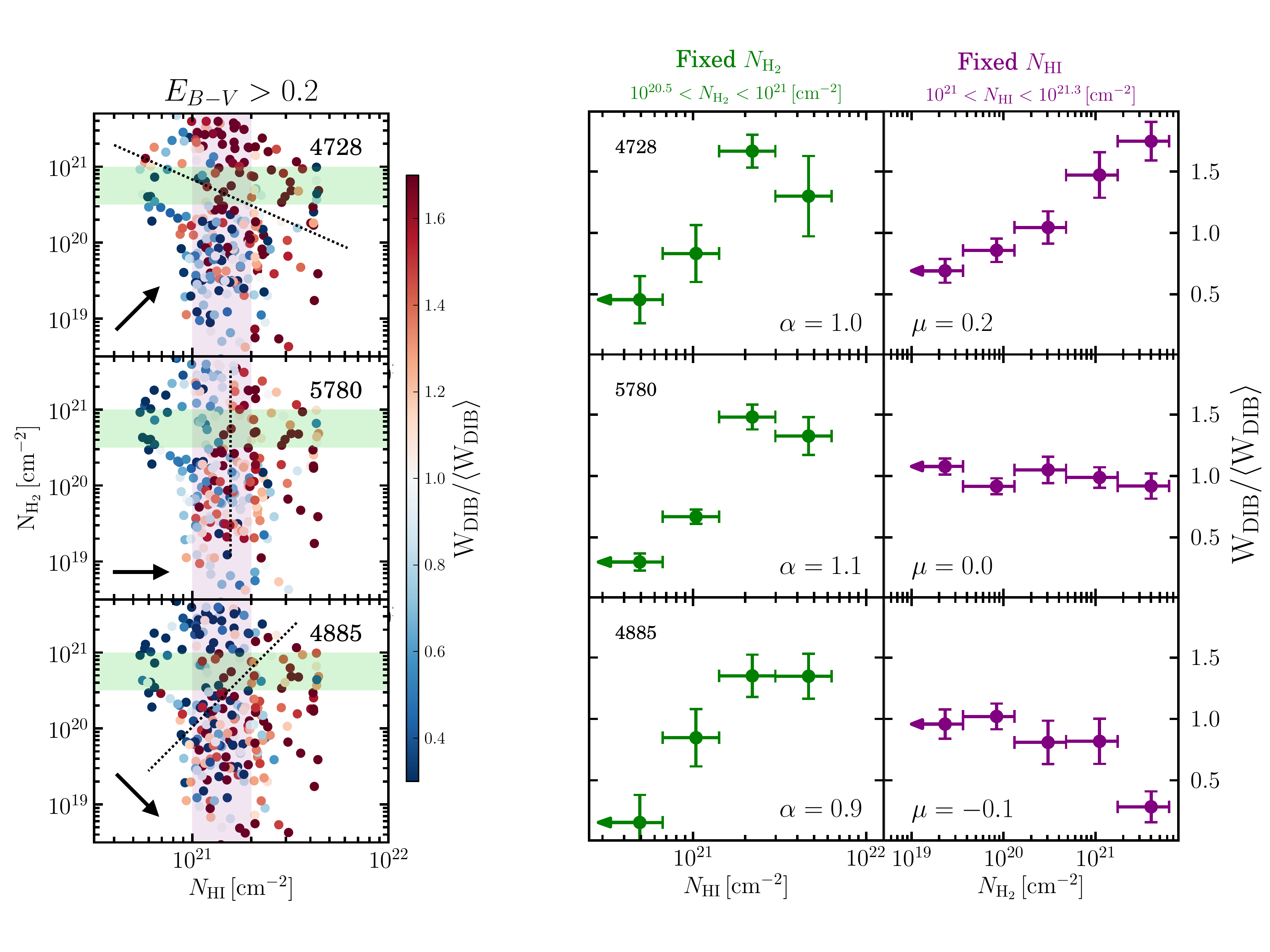}
\caption{Dependencies of absorption strength as a function of \Nhyi\ and \Nhytwo\ for three DIBs: $\lambda$4728, 5780 and 4885. The left panel shows the distribution of points in the \Nhyi\ , \Nhytwo\ plane for regions selected with $E_{B-V}>0.2$. To guide the eye, the dashed lines are inserted manually to separate the blue and red data points. The arrows indicate the direction of increasing DIB strength. The green and purple bands indicate the regions at fixed \Nhytwo\ and \Nhyi\ used in the right panels respectively.
\emph{Right}: median absorption strength as a function of \Nhyi\ (at fixed \Nhytwo) and \Nhytwo\ (at fixed \Nhyi). The decline of the DIB strength at the highest $\Nhyi$ in the middle panel is likely due to the contamination of $\Nhyi$ from the background given that most of lines of sight in the bin are towards the Galactic disc with $|b|<10^{\circ}$.
}
\label{plot:four_DIB_with_hydrogen_fraction}
\end{center}
\end{figure*}
%----------------------------------

%----------------------------------
% figure
%----------------------------------
\begin{figure*}[!t]
\begin{center}
\includegraphics[scale=0.40]{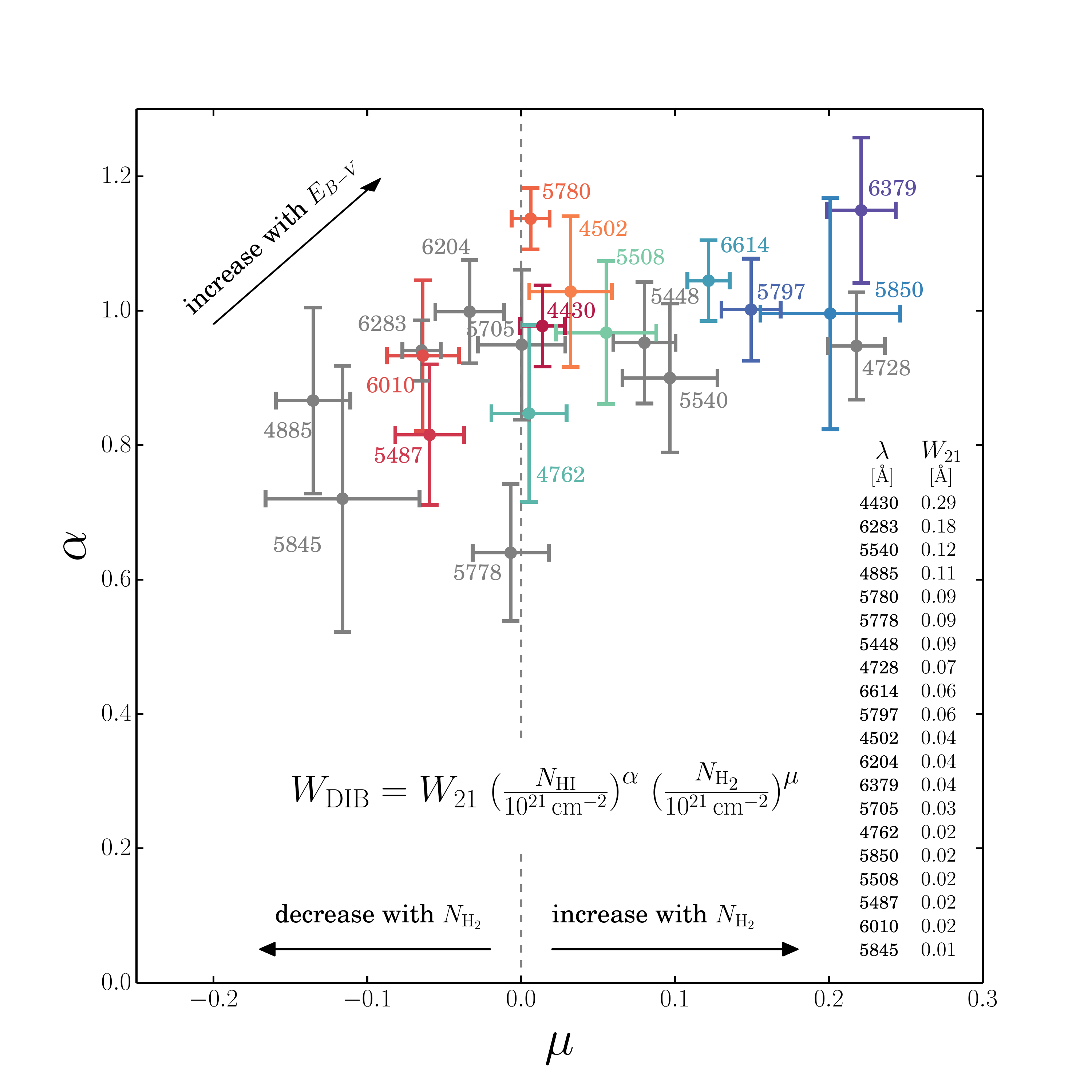}
\caption{Distribution of $\alpha$ and $\mu$ parameters characterizing the relation between DIB equivalent width, \Nhyi\ and \Nhytwo\ according to the relation indicated in the figure (equation~\ref{eq:main}). The values listed on the right show for each band the measured amplitude $W_{21}$. The colours indicate the line widths of DIBs from \citet{Hobbs2008}.  Narrow features, indicated with blue colours, appear to have preferentially positive $\mu$ values, while broad DIBs with red colours have $\mu\sim0$. DIBs blended with multiple lines are in grey. DIBs with $\mu>0$ are favoured in environments with higher molecular gas content.}
\label{fig:main}
\end{center}
\end{figure*}
%----------------------------------

%----------------------------------
% figure
%----------------------------------
\begin{figure*}[!t]
\begin{center}
\includegraphics[scale=0.4]{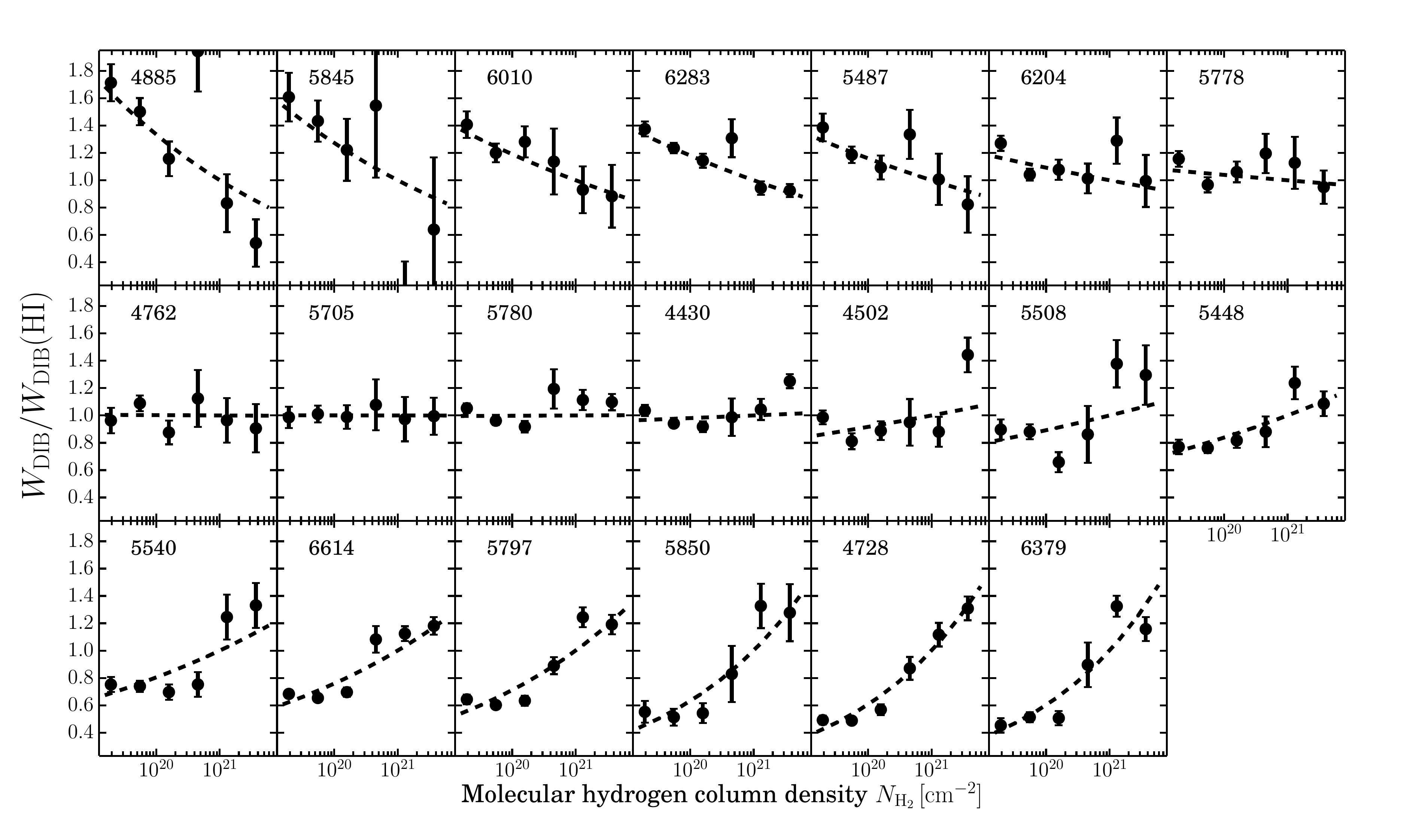}
\caption{DIB equivalent widths normalized by the dependences of atomic hydrogen, $W_{\rm DIB}(\rm HI)=W_{21}\;
\left(\frac{N_{\rm HI}}{10^{21}\,{\rm cm^{-2}}} \right)^{\alpha}$, as a function of the molecular hydrogen column densities. DIBs are ordered with their $\mu$ values from negative (top-left) to positive (bottom-right).  Data points are inverse variance-weighted mean of DIB equivalent width normalized by the dependences of atomic hydrogen and the error is estimated by bootstrapping. The dashed lines are the best-fitting dependences of molecular hydrogen, $\left(\frac{N_{\rm H2}}{10^{21}\,{\rm cm^{-2}}} \right)^{\mu}$.}
\label{fig:h2_dependences}
\end{center}
\end{figure*}

%----

%----------------------------------
% figure
%----------------------------------
\begin{figure*}[!ht]
\begin{center}
\includegraphics[scale=0.4]{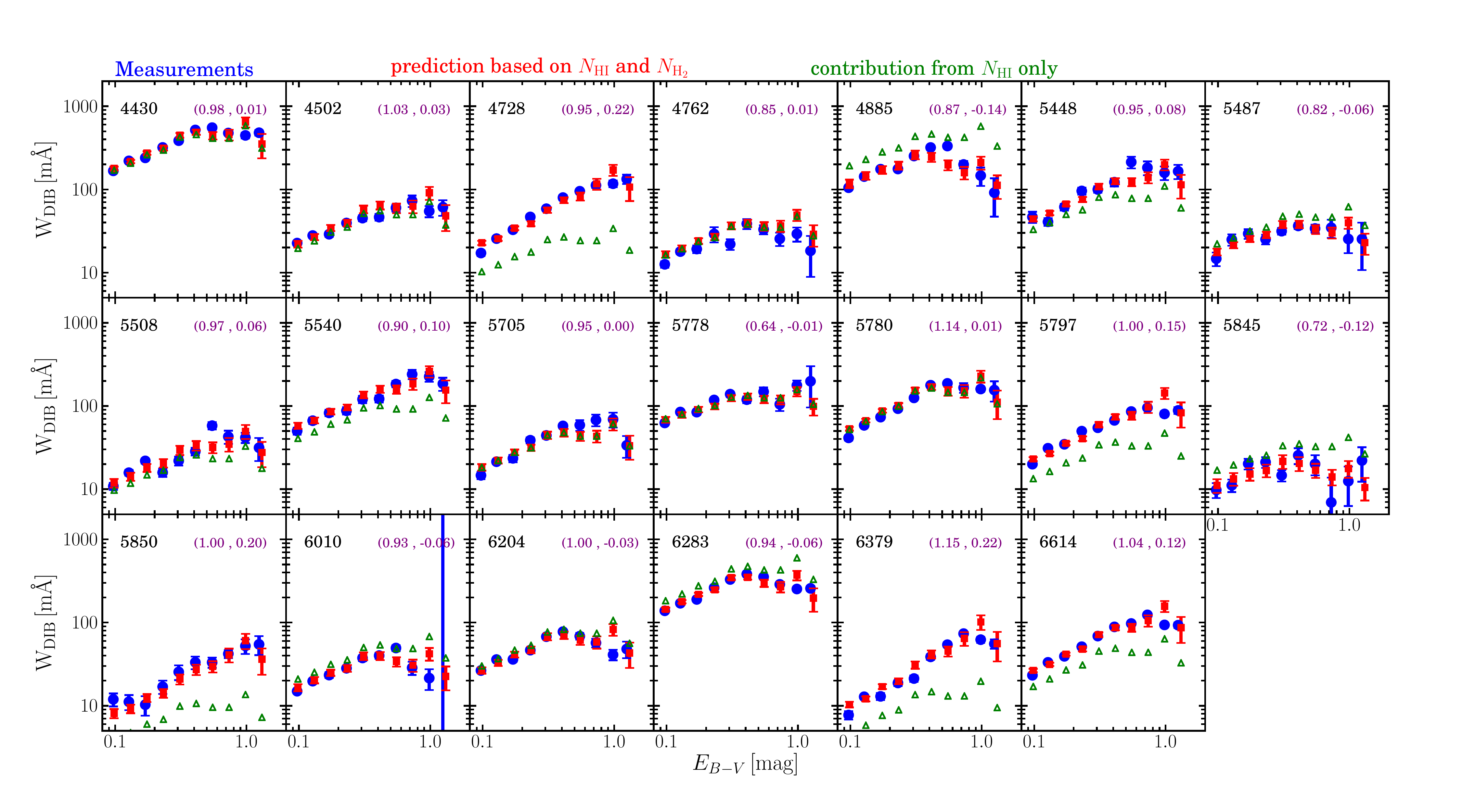}
\caption{Estimates of DIB absorption as a function of dust reddening based on our formalism involving only
(\Nhyi,\Nhytwo). The red squares are estimated equivalent widths using the median $\Nhyi$ and $\Nhytwo$ at each \ebv\ bin and equation~(\ref{eq:main}). Blue points show direct measurements. The green triangles are shown to illustrate the effect of \hytwo. To do so we show the estimated absorption strength only based on measured values of \Nhyi\ and fixing \Nhytwo$=10^{18}\,{\rm cm^{-2}}$. The purple numbers shown on the top-right corners show the ($\alpha$,$\mu$) values inferred for each DIB. 
}
\label{plot:f_h2_reproduction}
\end{center}
\end{figure*}
%----------------------------------

We now investigate the dependence of DIB absorption on the amount of hydrogen. The global rank-order correlation coefficient shown in Fig.~\ref{plot:DIB_vs_ISM_correlation} indicates a positive correlation between the strength of the total DIB absorption field and the amount of atomic hydrogen but a negative correlation with molecular hydrogen, traced by CO emission. As \Nhyi\ and \Nhytwo\ are not independent quantities at a fixed column density, the interpretation of the observed trends requires additional considerations.
Various authors have attempted to quantify correlations between DIB strength and atomic \emph{or} molecular hydrogen. However, it appears that only a few studies \citep[e.g.][]{Herbig1993} have investigated the dependence of DIBs on those two parameters \emph{simultaneously}.

In Fig.~\ref{plot:DIB_vs_H_H2} we show the relationships between the total DIB absorption, \Nhyi\ and \Nhytwo 
\footnote{We note that recent studies \citep[e.g.][]{Liszt2014a,Liszt2014b} have shown that the $N_{HI}$ and $E_{B-V}$ relation derived from emission measurements and the relation based on $Ly\alpha$ absorption lines and reddening in background stellar spectra are not fully consistent. We caution that such an effect needs to be considered when comparing the \emph{exact} values of derived parameters between DIBs and ISM column densities from emission-based measurements and from absorption- and reddening-based measurements. However, detailed investigations of systematics of these data sets are beyond the scope of this paper.}. 
Contrarily to what is observed with PAHs, we can see that at a fixed dust reddening value, the column density of \Nhyi\ and \Nhytwo\ affects the observed values of the total DIB equivalent width $W_{\rm all\,DIBs}$. This is illustrated in the inset of the figure which shows the DIB strength as a function of \Nhyi\ and \Nhytwo\ when selecting lines of sight with $0.2<$\ebv$<0.4\,$ mag. We find $W_{\rm all\,DIBs}$ to increase with \Nhyi\ but decrease with \Nhytwo. It is important to realize that \ebv\ can be used as a proxy for the total hydrogen column density. For example, \cite{Bohlin1978} derives that, on average, $N({\rm HI}+{\rm H_2})/E_{B-V}\simeq5.8\times 10^{21}\,{\rm atom\,cm^{-2}\,mag^{-1}}$, over a broad range of column densities. Selecting lines of sight within a narrow range of \ebv\ values therefore constrains the \emph{sum} of \Nhyi\ and \Nhytwo. 

We now investigate, for each of our 20 DIBs, the detailed relations between absorption strength, \Nhyi\ and \Nhytwo. To do so we first select regions of the sky with $\ebv>0.2$ mag to focus on the regime where different DIB behaviours are observed and select three DIBs ($\lambda4728$, $\lambda5780$ and $\lambda4885$)  representative of the range of correlations with molecular hydrogen. For each of them we show in Fig.~\ref{plot:four_DIB_with_hydrogen_fraction} the variation of their relative equivalent width as a function of both \Nhyi\ and \Nhytwo. The left panel shows the 2D distribution. The arrows show the directions in which the DIB absorption strength increases. We can observe that DIB $\lambda4728$ increases towards both higher \Nhyi\ and \Nhytwo, $\lambda5780$ increases only with \Nhyi, and $\lambda4885$ increases towards the lower right, where \Nhyi\ increases but \Nhytwo decreases. 
We note that the sampling of the \Nhyi-\Nhytwo\ space is not homogeneous and some care is needed to interpret details of the data point distribution
\footnote{We note that the sky sampling provided by the SDSS at high extinction, i.e. \ebv$\sim1\,$mag, originates mostly from a number of special fields with known molecular clouds. As a result, our sampling of high-dust extinction regions favors high \Nhytwo\ values, a selection effect which requires some care when interpreting measured correlations.}.
To show more clearly the above trends, for each of the three DIBs we select a narrow 
\Nhytwo\ bin, $10^{20.5}<N_{\rm H_2}<10^{21}\,{\rm cm^{-2}}$ corresponding to a region where \Nhyi\ spans a large range of value and where the sampling is not too inhomogeneous and measure the median DIB equivalent width as a function of \Nhyi\ , as shown in the right panel. We can see that, at fixed \Nhytwo, these three DIBs all become stronger with \Nhyi. The decline of the DIB strength at the highest $\Nhyi$ is likely due to the contamination of $\Nhyi$ from the background given that most of lines of sight in the bin are towards the Galactic disc with $|b|<10^{\circ}$.
Similarly, we select a narrow \Nhyi\ bin, $10^{21}<N_{\rm HI}<2\times10^{21}\,{\rm cm^{-2}}$ corresponding to a region where \Nhytwo spans a large range of value and where the sampling is not too inhomogeneous and show the median DIB equivalent width as a function of \Nhytwo . The different behaviour as a function of molecular hydrogen is clearly seen. At a given \Nhyi, DIB $\lambda4728$ is positively correlated with \Nhytwo, $\lambda5780$ is not affected by the presence of molecules, while $\lambda4885$ gets weaker at higher \Nhytwo.

%------------------------------------------------------
\begin{table}[!t] 
\caption{Best fit parameters characterizing the relation between DIBs and hydrogen column densities (equation~\ref{eq:main})}
%with $E_{B-V}>0.1$}
\centering
\begin{tabular}{cccc}
\hline\hline
$\lambda$ (DIB) & $\alpha$ (\hyi) & $\mu$ (\hytwo) & ${\rm W_{21}}$ \\ [0.5ex] % inserts table 
(\AA)  & -- & -- & (\AA)\\ [0.5ex]
\hline

4430 & $0.98 \pm 0.06$ & $+0.01 \pm 0.01$ & $0.287 \pm 0.012$  \\
4502 & $1.03 \pm 0.10$ & $+0.03 \pm 0.03$ & $0.038 \pm 0.003$  \\
$\ \, 4728^{*}$ & $0.95 \pm 0.09$ & $+0.22 \pm 0.02$ & $0.070 \pm 0.003$  \\
4762 & $0.85 \pm 0.12$ & $+0.01 \pm 0.03$ & $0.024 \pm 0.002$  \\
$\ \, 4885^{*}$ & $0.87 \pm 0.13$ & $-0.14 \pm 0.03$ & $0.111 \pm 0.011$  \\
$\ \, 5448^{*}$ & $0.95 \pm 0.09$ & $+0.08 \pm 0.02$ & $0.087 \pm 0.005$  \\
5487 & $0.82 \pm 0.12$ & $-0.06 \pm 0.03$ & $0.021 \pm 0.002$  \\
5508 & $0.97 \pm 0.12$ & $+0.06 \pm 0.04$ & $0.022 \pm 0.002$  \\
$\ \, 5540^{*}$ & $0.90 \pm 0.11$ & $+0.10 \pm 0.03$ & $0.118 \pm 0.008$  \\
$\ \, 5705^{*}$ & $0.95 \pm 0.11$ & $+0.00 \pm 0.03$ & $0.028 \pm 0.002$  \\
$\ \, 5778^{*}$ & $0.64 \pm 0.11$ & $-0.01 \pm 0.03$ & $0.089 \pm 0.007$  \\
5780 & $1.14 \pm 0.05$ & $+0.01 \pm 0.01$ & $0.089 \pm 0.004$  \\
5797 & $1.00 \pm 0.07$ & $+0.15 \pm 0.02$ & $0.058 \pm 0.002$  \\
$\ \, 5845^{*}$ & $0.72 \pm 0.21$ & $-0.12 \pm 0.05$ & $0.010 \pm 0.002$  \\
5850 & $1.00 \pm 0.15$ & $+0.20 \pm 0.04$ & $0.024 \pm 0.002$  \\
6010 & $0.93 \pm 0.13$ & $-0.06 \pm 0.02$ & $0.020 \pm 0.002$  \\
$\ \, 6204^{*}$ & $1.00 \pm 0.08$ & $-0.03 \pm 0.02$ & $0.037 \pm 0.003$  \\
$\ \, 6283^{*}$ & $0.94 \pm 0.05$ & $-0.06 \pm 0.01$ & $0.177 \pm 0.007$  \\
6379 & $1.15 \pm 0.11$ & $+0.22 \pm 0.02$ & $0.035 \pm 0.002$  \\
6614 & $1.04 \pm 0.06$ & $+0.12 \pm 0.01$ & $0.063 \pm 0.002$  \\
\hline
\vspace{.2cm}
\end{tabular} 
\begin{tablenotes}
\item *DIBs possibly blended with multiple weak DIBs
\end{tablenotes}
\label{table:NHI_NH2_dependence}
\end{table} 
%------------------------------------------------------

We now investigate the general behaviour of all of our 20 DIBs with atomic and molecular hydrogen. The previous examples, shown in Fig. \ref{plot:DIB_vs_H_H2} and Fig. \ref{plot:four_DIB_with_hydrogen_fraction}, motivate a formulation of the overall DIB equivalent width dependence as power-law functions of \Nhyi\ and \Nhytwo:
\begin{equation}
\hat W_{\rm DIB} = 
W_{21}\;
\left(\frac{N_{\rm HI}}{10^{21}\,{\rm cm^{-2}}} \right)^{\alpha}\;
\left(\frac{N_{\rm H_{2}}}{10^{21}\,{\rm cm^{-2}}} \right)^{\mu}\;,
\label{eq:main}
\end{equation}
where $W_{21}$ is a normalization denoting the relative strength of each absorption feature (similar to $W/A_V$). We estimate the three parameters of the relation,$W_{21},\alpha$ and $\mu$, by minimizing a global $\chi^2$: 
\begin{equation}
\chi^2 = 
\sum_i
\frac{ \left(W_{{\rm DIB},i} - \hat W_{\rm DIB}(W_{21},\alpha,\mu) \right)^2}
{\sigma_{W,i}^2}\;,
\end{equation}
where $W_{{\rm DIB},i}$ is the observed equivalent width of a DIB in a given pixel $i$ of the sky map. 
Applying directly this $\chi^2$ estimate to the entire data set is subject to sampling and selection effects.
Since the vast majority of the lines of sight probe low column density values (see Fig.~\ref{plot:DIB_vs_H_H2}), a straight $\chi^2$ evaluation would effectively only have constraining power at low \Nhyi\ values. In addition, our sampling of the high-end edge of the \Nhyi--\Nhytwo\ space is relatively poor.
To select a more homogeneously sampled space and reduce the effect of overestimating $\Nhyi$ towards the Galactic disc, we only consider lines-of-sight with \ebv$>0.1\,{\rm mag}$, $\Nhyi<10^{21.3}\,{\rm cm^{-2}}$ and $\rm W_{CO10} > 0 \, K\,km/s$. This allows us to estimate the parameters more robustly, at the cost of losing a fraction of the data set.
We note that the values of the best-fitting parameters $\alpha$ and $\mu$ are likely to be biased due to the use of emission-based column density estimations.  
In addition, the values also vary depending on the range of Galactic reddening and hydrogen column density values we select. 
However, in Appendix~\ref{appendix_c} we show that the relative distances between pairs of the 20 ($\alpha$,$\mu$) points do not change appreciably with different selections. In the following, we will only focus on their relative distances.

The results of the fitting are presented in Table~\ref{table:NHI_NH2_dependence}. The errors are obtained by bootstrapping the sample $200$ times. Fig.~\ref{fig:main} shows the distribution of values as a function of $\alpha$ and $\mu$. The colour represents the line width of each DIB measured by \citet{Hobbs2008} with bluer indicating narrower profile. DIBs which are potentially blended with multiple components are in grey colour. We find that \emph{all} of the 20 DIBs show positive correlations with the amount of atomic hydrogen. The mean DIB equivalent widths are found to scale like $\Nhyi^\alpha$ with $\alpha\sim1$. In contrast, we find a range of correlations with molecular hydrogen: $W_{\rm DIB}\propto\Nhytwo^\mu$ with $-0.2\lesssim\mu\lesssim0.2$. Certain DIBs, such as $\lambda$5780, $\lambda$4430, $\lambda$4762 have $\mu$ values consistent with zero and therefore are not sensitive to the amount of molecular hydrogen. 
Fig.~\ref{fig:h2_dependences} shows the inverse variance-weighted mean of DIB equivalent widths normalized by the dependences of atomic hydrogen, $W_{\rm {DIB}}(\rm {HI})=W_{21}\;
\left(\frac{N_{\rm HI}}{10^{21}\,{\rm cm^{-2}}} \right)^{\alpha}$, as a function of molecular hydrogen column densities. DIBs are ordered by their $\mu$ values from negative (top-left) to positive (bottom-right). The figure shows that after removing the dependences on the $\Nhyi$\ ,  the mean strength of DIBs with $\mu<0$ decreases with \Nhytwo\ while for those with $\mu>0$ strengthens. We also map the relative strength of DIBs with different $\mu$ values in the sky \footnote{The maps of DIBs having different correlations with molecules can be viewed at \url{http://www.pha.jhu.edu/~tlan/DIB_SDSS/}.}. The map of DIBs with $\mu > 0$ appears to have more weight towards molecular clouds. We note that given the lower absolute values of $\mu$ compared to $\alpha$, we expect \Nhyi\ to be the main parameter characterizing DIB equivalent widths and $\mu$ to be of secondary importance. The distribution of $(\alpha,\mu)$ values does not appear to be correlated with the relative strength of the DIBs, $W_{21}$. We observe a tendency for narrower DIBs to have higher $\mu$ values, which is also shown in \citet{Welty2013}.

A number of interesting implications derive from the observed distribution of the $\alpha$ and $\mu$ values:
\begin{itemize}
\item It is expected that if two DIBs are formed from transitions between a single ground state and two different vibronic levels, their measured strengths should be perfectly correlated, with a correlation coefficient of unity. DIBs with different values of $(\alpha,\mu)$ cannot correlate perfectly with each other and are likely due to different carriers. Conversely, DIBs for which measured values of $(\alpha,\mu)$ are consistent with each other \emph{may} originate from the same carrier or belong to a same `family'. Note that this statement depends on the accuracy with which the $(\alpha,\mu)$ parameters can be measured. Our analysis does not reveal any clustering of the points in the $(\alpha,\mu)$ plane. Our measurements suggest a continuum in the properties of each of the 20 selected bands rather than the existence of a few families.
\item At a fixed value of \Nhyi, \Nhytwo\ can vary by orders of magnitude. The \hytwo\ dependence, which exists when $\mu\neq0$, explains a large fraction of the scatter observed in measurements of $W_{\rm DIB}(N_{\rm HI})$ and $W_{\rm DIB}(E_{B-V})$.
\item Since dust column density is known to be roughly proportional to the \emph{total} amount of hydrogen, we expect DIBs with high $\alpha$ and $\mu$ values to correlate more strongly with dust reddening. 
Such a trend is observed in our analysis. DIB $\lambda6379$, $\lambda4728$ and $\lambda5850$ with higher $\alpha$ and $\mu$ also have steeper slopes $\gamma$ with $E_{B-V}$ among the 20 DIBs shown in Table~\ref{table:best_fit_power_law}.
\item DIBs with different $\mu$ values will behave differently in environments with higher molecular fraction. DIB line ratios $W(\lambda_1)/W(\lambda_2)$ will be correlated with $\mu(\lambda_1)-\mu(\lambda_2)$. This provides us with a generalization of the $\sigma,\zeta$ effect discussed in the literature (see Section~\ref{sec:discussion}). The distribution of points in Fig.~\ref{fig:main} can be used to predict that larger line ratios are expected for pairs of DIBs with greater $\Delta\mu$, for example between DIBs $\lambda$4728 and $\lambda$4885, when probing lines of sight with higher molecular fractions.
\item As shown in equation~(\ref{eq:main}), the mean equivalent width of a DIB can be parametrized as a function of both \Nhyi\ and \Nhytwo\ . This implies that the knowledge of the equivalent widths for two or more DIBs with different $(\alpha,\mu)$ values can be used to infer, statistically, both \Nhyi\ and \Nhytwo\ along the corresponding lines of sight.
\item We note that the dependence between $W_{\rm DIB}$ and (\Nhyi, \Nhytwo) can be used to predict the relation between DIB equivalent width and other ISM tracers. This can be used to explore whether additional parameters are important in describing DIB behaviours.
\end{itemize}

Here we illustrate the above point by attempting to reproduce the trends observed between $W_{\rm DIB}$ and \ebv. To do so we use the same sampling of the sky as done in Section~\ref{sec:reddening}. For each \ebv\ bin we estimate the median values of \Nhyi\ and \Nhytwo\ and estimated $W_{\rm DIB}(E_{B-V})$ using equation~(\ref{eq:main}) and without any knowledge of \ebv. The results are shown in Fig.~\ref{plot:f_h2_reproduction} using red points. For comparison, we show the direct reddening-based measurements presented in Section~\ref{sec:reddening} in blue. As can be seen, our hydrogen-based formalism provides us with a good description of all the trends given by the data -- for 20 DIBs over more than an order of magnitude in \ebv. In particular it naturally reproduces the turnover at the high end seen for only specific DIBs, the so-called skin effect. This shows that the anti-correlation between DIB strength and dust in the high column density regime can be explained by the correlation between hydrogen and DIBs only.

To further illustrate the meaning of the $(\alpha,\mu)$ parameters we show expected DIB equivalent widths considering only the \hyi\ dependence. To do so we fix the amount of \hytwo\ to a low value: $\Nhytwo=10^{18}\,{\rm cm^{-2}}$. The results are shown with the green triangles. For about a third of the DIBs, we see that \hytwo\ has virtually no effect on the relation between $W$ and \ebv. This is the case for DIBs with $\mu\sim0$. Similarly we can see the level at which molecules can influence the trends seen for DIBs with $\mu\neq 0$.

Describing the behaviour of DIBs using only \hyi\ and \hytwo\ allows us to describe a wide range of observed properties, i.e. relations between DIBs themselves, between mean DIB strength and ISM tracers as well as some of the related scatter observed in the corresponding distributions. We note that equation~(\ref{eq:main}) is not expected to reproduce all observed behaviours. Additional dimensions might be required. However our approach has shown that a large fraction of the DIB dependencies and variances could be simply explained by their relations to atomic and molecular hydrogen.

% % % % % % % % % % % % % % % % % % % % 
\section{Discussion}
\label{sec:discussion}
% % % % % % % % % % % % % % % % % % % % 

Our analysis has allowed us to map out the strength of 20 DIBs over a large fraction of the sky and derive a simple formulation of the mean DIB equivalent width as a function of only atomic and molecular hydrogen column densities.
So far, most DIB analyses focused on \emph{projections} of the relation $W_{\rm DIB}=f(N_{\rm HI}, \Nhytwo)$. Only a few \citep{Herbig1993,Thorburn2003} considered the simultaneous dependence on multiple ISM tracers but were based on small samples.
As most interstellar quantities show a positive correlation with each other, simply due to the increase of interstellar material with distance, pairwise correlations between observables are often driven by this effect. This is most clearly reflected in the correlation of DIBs with extinction. A multi-dimensional approach is required to disentangle different effects. We now show that our results, and in particular the distribution of $(\alpha,\mu)$ values presented in Fig.~\ref{fig:main} and Table~\ref{table:NHI_NH2_dependence}, are consistent with numerous observational results reported in the literature, shed light on the origin of various correlations, and can be used to predict correlations not yet measured.

%------------------------------------------------------
\subsection{DIB families}

Various authors have studied cross-correlations between the strength of different DIBs or compared their dependences on ISM tracers to assess whether or not they may belong to families. If not, they must arise from different carriers. Similarly, with our formalism, DIBs with different values of $(\alpha,\mu)$ cannot correlate perfectly with each other and are likely due to different carriers.

Investigating different dependences with dust reddening, \citet{Josafatsson1987} studied six DIBs and defined two classes: the first one consists of DIBs $\lambda$5780, $\lambda$5797 and $\lambda$5850 and the second group contains the broader bands $\lambda$5778 and $\lambda$5845. This is consistent with our decomposition. DIB $\lambda$5778 and $\lambda$5845 are on the lower left side on the figure and are expected to correlate less with dust. The $(\alpha,\mu)$ values of the two group are inconsistent with each other.

\citet{Cami1997} studied 44 DIBs and identified two families based on the degree of correlation between bands.
The first one contains DIBs $\lambda$5797, $\lambda$6379 and $\lambda$6614 and the second one the $\lambda$4502, $\lambda$5789, $\lambda$6353 and $\lambda$6792 DIBs. For the DIBs in common with our analysis, this decomposition is consistent: the members of the first group all live in the upper right corner of Fig.\ref{fig:main} have high $\alpha$ and high $\mu$ values. DIB $\lambda$4502 which belongs to the second group has a lower $\mu$ value. Reported correlations between additional DIBs and these two families appear to be in agreement with our findings.

\citet{Friedman2011} investigated cross-correlations between DIB strengths for a selection of eight bands: they report that the two weakest pairwise correlations are found from ($\lambda$5797, $\lambda$6283) and ($\lambda$5797, $\lambda$5487). These two pairs correspond to the two largest distances in our $(\alpha,\mu)$ plane. They also report that the correlation of DIBs $\lambda$5780 and $\lambda$5705 is high. We note that the $(\alpha,\mu)$ values of these two bands are consistent with each other.

As mentioned in \citet{Cox2005} the literature indicates the existence of families of DIBs based on the shape of their profile \citep[e.g.][]{Josafatsson1987,Krelowski1987,Cami1997,Porceddu199,Moutou1999,Wzsolek2003,Galazutdinov2003}. Members of one group have narrow profiles with a sub-structure that is indicative of a gas-phase molecule \citep{Sarre1995,Ehrenfreund1996}. They are DIBs $\lambda$5797, $\lambda$5850, $\lambda$6196, $\lambda$6379 and $\lambda$6614. We note that all of them appear clustered in the upper right hand side of the $\alpha,\mu$ plane. A second group, with DIBs $\lambda$5780, $\lambda$6283 and $\lambda$6204, has absorption features with no apparent substructure \citep[\eg][]{Cami1997}. We note that all these DIBs have $\mu\sim0$.

%------------------------------------------------------
\subsection{Correlations with atomic hydrogen}

Correlations between DIBs and the amount of atomic hydrogen have been reported for a long time \citep{Herbig1993}. In their analysis of eight DIBs, \citet{Friedman2011} reported cross-correlation coefficients between \Nhyi\ and eight DIBs: $\lambda$5780, $\lambda$6204, $\lambda$6283, $\lambda$6196, $\lambda$6614, $\lambda$5705, $\lambda$5797, and $\lambda$5487, in decreasing order of correlation amplitude. This trend is consistent with our results within the uncertainties: $\lambda$5780 \& $\lambda$6204 have $\mu\sim0$ while $\lambda$5797 and $\lambda$5487 have $\mu\sim0.1$ and $-0.05$, respectively. The fact that DIB $\lambda$5780 has the highest degree of correlation with hydrogen is reflected by the fact that it has the highest $\alpha$ value and $\mu$ consistent with zero. 
\citet{Welty2013} also found that DIB$\lambda$5780 tends to be weaker with higher molecular \emph{fraction} (less atomic hydrogen) which is consistent with our results.

%------------------------------------------------------
\subsection{Molecules and the $\sigma,\zeta$ dichotomy}

\begin{figure}[t]
\begin{center}
\includegraphics[scale=0.5]{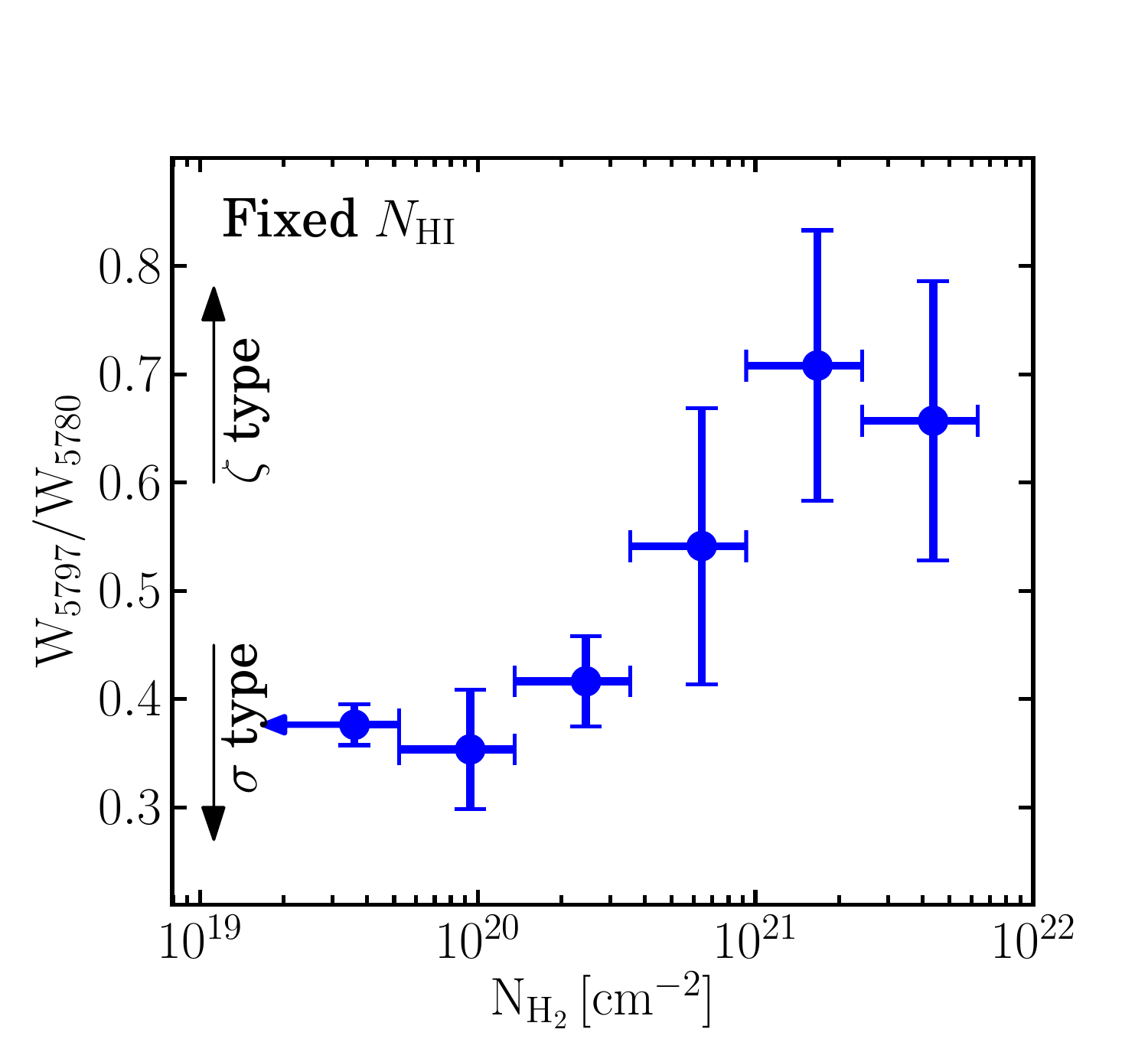}
\caption{The ratio between the equivalent width of DIB $\lambda$5797 and DIB$\lambda$5780 as a function of molecular hydrogen column density for lines of sight selected with $10^{21}<N_{\rm HI}<10^{21.3}\,{\rm cm^{-2}}$.}
\label{plot:ratio}
\end{center}
\end{figure}

\citet{Krelowski1988} found that the $\lambda5797$/$\lambda5780$ ratio is higher towards star $\zeta$ Oph than towards $\sigma$ Sco. As molecules (\eg CN, CH) are more abundant towards $\zeta$ Oph \citep[][]{Danks1984}, it was suggested that clouds with higher molecular fractions cause stronger $\lambda5797$ absorption than $\lambda5780$. This result introduced the $\sigma/\zeta$ types of lines of sight. 
We can illustrate this effect with our measurements. In Fig.~\ref{plot:ratio} we show the measured ratio between the equivalent width of DIBs $\lambda$5797 and $\lambda$5780, as a function of molecular hydrogen column density for lines of sight selected within a narrow range of atomic hydrogen column densities: $10^{21}<N_{\rm HI}<2\times10^{21}\,{\rm cm^{-2}}$. As clearly seen, the ratio between these two lines increases with higher molecular fraction or in other words when transitioning from $\sigma$-type lines of sight to $\zeta$-type.

%----------------------------------
% figure
%----------------------------------
%----------------------------------

\citet{Kos2013_family} observed 19 DIBs and classified them into two groups according to their correlations with dust as a function of the $\sigma/\zeta$ type sightlines. Their classification is consistent with the results derived with our formalism. DIBs such as $\lambda$5780, $\lambda$5705 and $\lambda$6204 having values of $\mu$ consistent with zero behave similarly and are classified into the same group (Type I), while DIBs $\lambda$6614, $\lambda$5797, $\lambda$5850 and $\lambda$6379 living in the right hand side of Fig.~\ref{fig:main} with $\mu\simeq0.1-0.2$ are classified into the other group (Type II).

In their analysis of eight DIBs, \citet{Friedman2011} reported that DIBs $\lambda$6614 and $\lambda$5797 have the strongest correlation coefficients with \hytwo. This is consistent with the fact that, among their selected DIBs, those two have the highest $\mu$ values.

Our findings are also consistent with the results of \citet{Thorburn2003}. These authors found DIB$\lambda4728$ correlates with molecules traced by $C_2$. In addition, they found no correlation between $W(\lambda6614)$/$W(\lambda6196)$ and molecules (traced by $C_2$, CN and CH) and anti-correlation $W(\lambda6204)$/$W(\lambda6196)$. This is in agreement with the fact that $\mu(\lambda6614)>\mu(\lambda6204)$.
Similarly, \citet{Krelowski1999} and  \citet{Vos2011} showed that  $W(\lambda5797)$/$W(\lambda5780)$ is positively correlated with $W({\rm CH})/E_{B-V}$. This is in line with the fact that $\mu(\lambda5797)>\mu(\lambda5780)$. Interestingly the ratio $W(\lambda5797)$/$W(\lambda5780)$ does not correlate with $W({\rm CH}^+)/E_{B-V}$. %This may indicate that the level of correlation with molecules characterized by $\mu$ only applies to neutral species.

%------------------------------------------------------
\subsection{Correlations with dust}

As pointed out earlier, dust reddening can be used as a proxy for the total hydrogen column density:
$N({\rm HI}+{\rm H_2})/E_{B-V}\simeq5.8\times 10^{21}\,{\rm atom\,cm^{-2}\,mag^{-1}}$ \citep{Bohlin1978}. In Fig.~\ref{plot:f_h2_reproduction} we have already shown that it is possible to reproduce the trends observed between the mean DIB equivalent widths and \ebv\ by considering only the amount of \hyi\ and \hytwo\ present along the lines-of-sight. The break seen at high \ebv\ values, also called the skin effect \citep{Snow1974}, can be quantitatively characterized by the decrease of \hyi\ in dense clouds, with higher molecular fraction. As indicated in Fig.~\ref{plot:DIB_vs_H_H2} some of the scatter in the relation between $W_{\rm DIB}$ and dust reddening originates from the existence of a distribution of \Nhyi\ and \Nhytwo\ values along the corresponding lines of sight.\\

Finally, we note that our ability to reproduce a large range of observational results using the $(\alpha,\mu)$ parametrization derived in Section~\ref{sec:DIB_and_f_h2} indicates that the statistical measurements of DIB equivalent widths derived from the analysis of the SDSS are robust. The existence of a relation between $W_{DIB}$, \Nhyi\ and \Nhytwo\ explains a large fraction of the scatter observed when considering correlations between the equivalent width of a DIB and \emph{one} ISM tracer, such as \Nhyi, \Nhytwo, molecular fraction or \ebv.\\

\section{Summary}

We have used about $500,000$ spectra of stars, quasars and galaxies taken by the SDSS to map out the distribution of DIBs induced by the ISM of the Milky Way. We have showed that, after carefully removing the intrinsic SED of each source and taking care of spectroscopic calibration effects and spectral features due to the Earth's atmostphere, it is possible to detect absorption features at the $10^{-3}$ level. This allows us to detect more than 20 DIBs from $\lambda=4400$ to $6700\,{\rm \AA}$ and measure their strength as a function of position on the sky. Focusing on a set of 20 bands, for which we can robustly characterize the line properties, we have created a map of DIB absorption covering about 5,000 deg$^{2}$. This map can be used to measure correlations with various tracers of the ISM: atomic and molecular hydrogen, dust and PAHs. Our findings can be summarized as follows:
\begin{itemize}

\item For each of the 20 selected DIBs, we have measured their mean absorption as a function of dust reddening and found results consistent with previous studies based on individual spectra of hot stars. For certain DIBs, we observe a break at high \ebv\ value, above which the absorption strength decreases with dust column density, the so-called skin effect. 
%We find similar results when correlating DIBs with PAHs.
%

\item As various ISM tracers are correlated with each other, a multi-dimensional approach is required to disentangle different effects. Investigating the dependence of DIB absorption strength on atomic and molecular hydrogen simultaneously we find that, on average, the three quantities can be described by 
\begin{equation}
W_{\rm DIB} \propto (N_{\rm HI})^\alpha \, (N_{\rm H_2})^\mu\, \nonumber
\end{equation}
(see equation~4). For all DIBs we find $\alpha\sim1$ but we find a range of values for $\mu$, from $-0.2$ to $+0.2$, indicating that different DIBs have a different affinity to molecules. DIBs with $\mu>0$ are favoured in environments with higher molecular gas content. This parametrization also shows that \Nhyi\ is the main parameter characterizing DIB equivalent widths. The effect of \Nhytwo\ is, in general, of secondary importance but can dominate in certain regimes for DIBs with $\mu$ values departing from zero.

\item We show that the combined dependence on both \Nhyi\ and \Nhytwo\ can be used to reproduce a number of observational results. For example, one can reproduce the observed trends between DIB absorption and dust reddening (including the so-called skin effect) using only the hydrogen-based parametrization. We also note that the combined dependence on \Nhyi\ and \Nhytwo\ can largely explain the scatter observed in a number of correlations between DIBs and ISM tracers.

\item We estimate the $(\alpha,\mu)$ values for the 20 DIBs and study their distribution. DIBs with different $(\alpha,\mu)$ values cannot correlate perfectly with each other and are likely due to different carriers.
We show that the inferred $(\alpha,\mu)$ values are consistent with numerous observational results reported in the literature and shed light on the origin of various correlations: (i) the relations between DIBs themselves; (ii) relations between DIB strength and ISM tracers; (iii) the $\sigma$/$\zeta$ dichotomy, which can be generalized to any pair of DIBs; (iv) and some of the related scatter observed in the corresponding distributions. The estimated $(\alpha,\mu)$ values can also be used to predict correlations not yet measured. We note that the consistency with such a broad set of observational results demonstrates the robustness of the equivalent width measurements of the selected 20 DIBs from SDSS stellar spectra.

\end{itemize}

While the origin of DIBs is still a mystery, our approach provides us with a new view and parametrization of numerous observational results previously reported and a metric to characterize the affinity between DIBs. Our parametrization of the strength of DIB absorption as a function of atomic and molecular hydrogen column densities might help us shed light on the physical mechanisms involved with the production and destruction of the DIB carriers.

\acknowledgements
We would like to thank Daniel Welty for his comments on an earlier version of the paper and his constructive report at the refereeing stage. We thank Tomaz Zwitter, Jacek Krelowski, and Don York for their comments on the manuscript. We also thank David Neufeld and Scott Friedman for useful discussions and James Gunn and Murdock Hart for providing the atmospheric absorption and emission spectra. This work was supported by NSF Grant AST-1109665, the Alfred P. Sloan foundation and a grant from Theodore Dunham, Jr., Grant of Fund for Astrophysical Research. G.Z. acknowledges partial support for this work provided by NASA through Hubble Fellowship grant \#HST-HF2-51351.001-A awarded by the Space Telescope Science Institute, which is operated by the Association of Universities for Research in Astronomy, Inc., for NASA, under contract NAS 5-26555

Funding for the SDSS and SDSS-II has been provided by the Alfred P. Sloan Foundation, the Participating Institutions, the National Science Foundation, the U.S. Department of Energy, the National Aeronautics and Space Administration, the Japanese Monbukagakusho, the Max Planck Society, and the Higher Education Funding Council for England. The SDSS Web Site is http://www.sdss.org/. 

Funding for SDSS-III has been provided by the Alfred P. Sloan Foundation, the Participating Institutions, the National Science Foundation, and the U.S. Department of Energy Office of Science. The SDSS-III web site is http://www.sdss3.org/.

SDSS-III is managed by the Astrophysical Research Consortium for the Participating Institutions of the SDSS-III Collaboration including the University of Arizona, the Brazilian Participation Group, Brookhaven National Laboratory, Carnegie Mellon University, University of Florida, the French Participation Group, the German Participation Group, Harvard University, the Instituto de Astrofisica de Canarias, the Michigan State/Notre Dame/JINA Participation Group, Johns Hopkins University, Lawrence Berkeley National Laboratory, Max Planck Institute for Astrophysics, Max Planck Institute for Extraterrestrial Physics, New Mexico State University, New York University, Ohio State University, Pennsylvania State University, University of Portsmouth, Princeton University, the Spanish Participation Group, University of Tokyo, University of Utah, Vanderbilt University, University of Virginia, University of Washington, and Yale University.

This publication makes use of data products from the Wide-field Infrared Survey Explorer, which is a joint project of the University of California, Los Angeles, and the Jet Propulsion Laboratory/California Institute of Technology, funded by the National Aeronautics and Space Administration.

Based on observations obtained with Planck (http://www.esa.int/Planck), an ESA science mission with instruments and contributions directly funded by ESA Member States, NASA, and Canada.

\appendix

\section{SDSS spectral calibration residuals}
\label{appendix_a}
\begin{figure*}
\center
\includegraphics[scale=0.4]{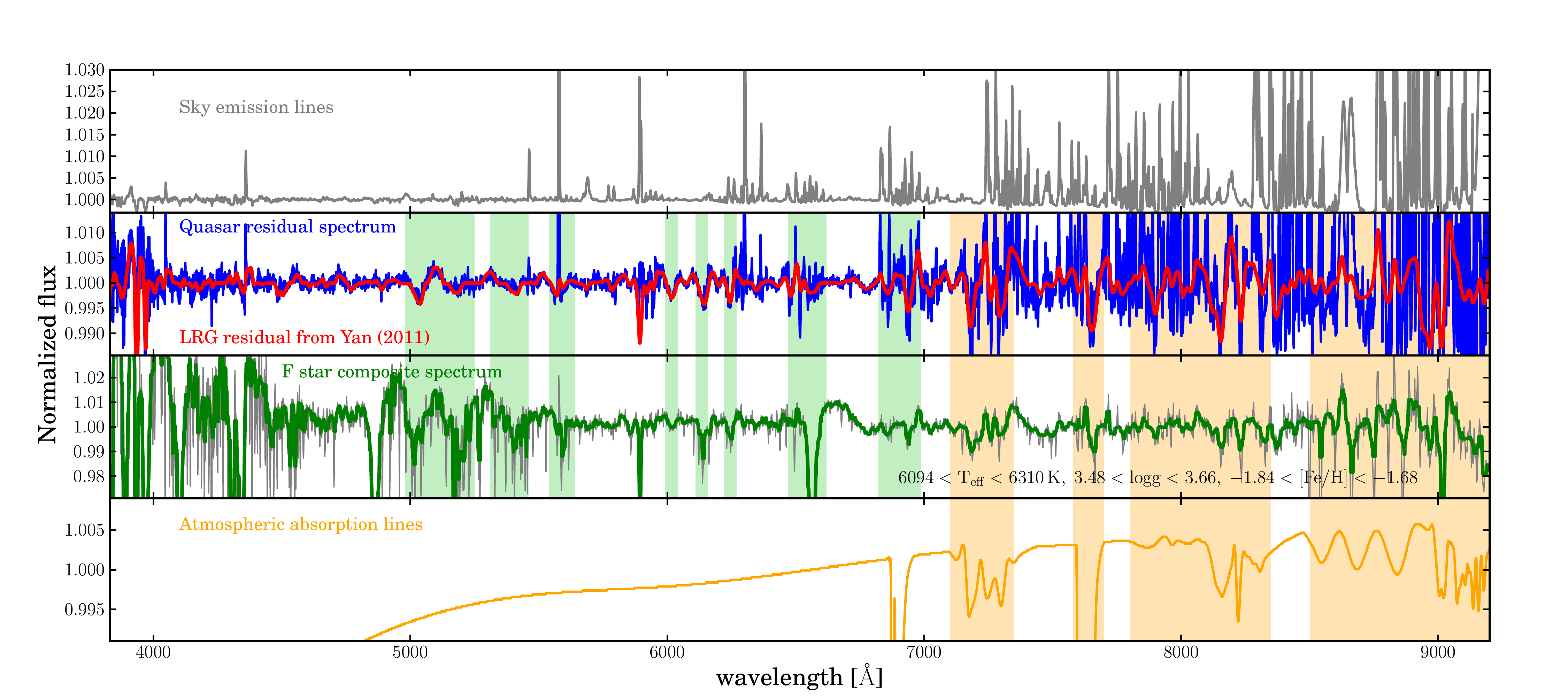}
\caption{Effects of atmospheric emission and absorption lines and the spectral calibration on the SDSS spectra. From top to bottom, we show the composite sky emission line spectrum (grey), the composite quasar residual spectrum (blue), the composite residual spectrum of luminous red galaxies from \citet[][red]{Yan2011}, a F-star composite spectrum (green), and an atmospheric absorption line spectrum (orange). We indicate features due to atmospheric absorption lines with orange vertical bands.  At $\lambda<7000\,$\AA, we find that many features correspond to the intrinsic stellar absorption features. Several distinct features are indicated with green vertical bands.}
\label{plot:small_fluc}
\end{figure*}

We investigate the nature of the imperfections of the spectral calibration of SDSS, as indicated by the residuals shown in Fig.~2. To investigate the origin(s) of such features, we compare the residuals with the absorption lines caused by the atmosphere and the absorption features intrinsic to F stars that are used in the SDSS spectral calibration. In Fig.~\ref{plot:small_fluc}, we show the composite residual spectra of quasars (blue) at \ebv$<0.02$ mag within the whole wavelength range covered by SDSS. In addition, we also show a composite spectrum of the SDSS sky fibers in grey (M. Hart, private communication), i.e., fibers that were pointed to the fields where there is no detectable source in imaging, and the composite residual spectrum of luminous red galaxies by \citet[][red]{Yan2011}. In addition, we show a measurement of an atmospheric absorption spectrum (J. Gunn, private communication) in orange and an F-star composite spectrum from our data-driven stellar model in green.

We find that, beyond around $7000\,$\AA, the composite residual spectra of quasars and galaxies are dominated by the residuals of atmospheric {\it emission} lines, mostly due to OH and ${\rm H_2O}$. However, the atmospheric {\it absorption} lines, which were {\it not} included in the calibration, also imprint absorption features in the final residuals. The residuals of sky emission lines and absorption lines prevent us from investigating DIBs in the red end of the wavelength coverage.

At $\lambda<7000\,$\AA, we find several features with no corresponding sky emission and absorption lines. To calibrate the flux of sources in each field, SDSS used standard F stars\footnote{\tt{http://www.sdss3.org/dr8/algorithms/spectrophotometry.php}}, selected based on photometric colours. The green line shows a typical F-star template normalized to unity using a running median filter. We see many of the residuals have corresponding features in the intrinsic F-star SED. Note this F-star template is a composite spectrum of F-stars observed in SDSS, so residuals due to atmospheric absorption lines are also evident.

These systematic residuals exist in all the SDSS spectra and extreme care must be taken while studying small spectral features, especially in the observer frame. In our analysis, we use sources at low-extinction regions as references and empirically remove these features.

%----------------------------------
% figure
%----------------------------------
%\begin{sidewaysfigure}
%\end{sidewaysfigure}
%----------------------------------

\section{DIB-dust correlations with different \ebv\ estimation}
\label{appendix_b}
\begin{figure*}[!ht]
\begin{center}
\includegraphics[scale=0.42]{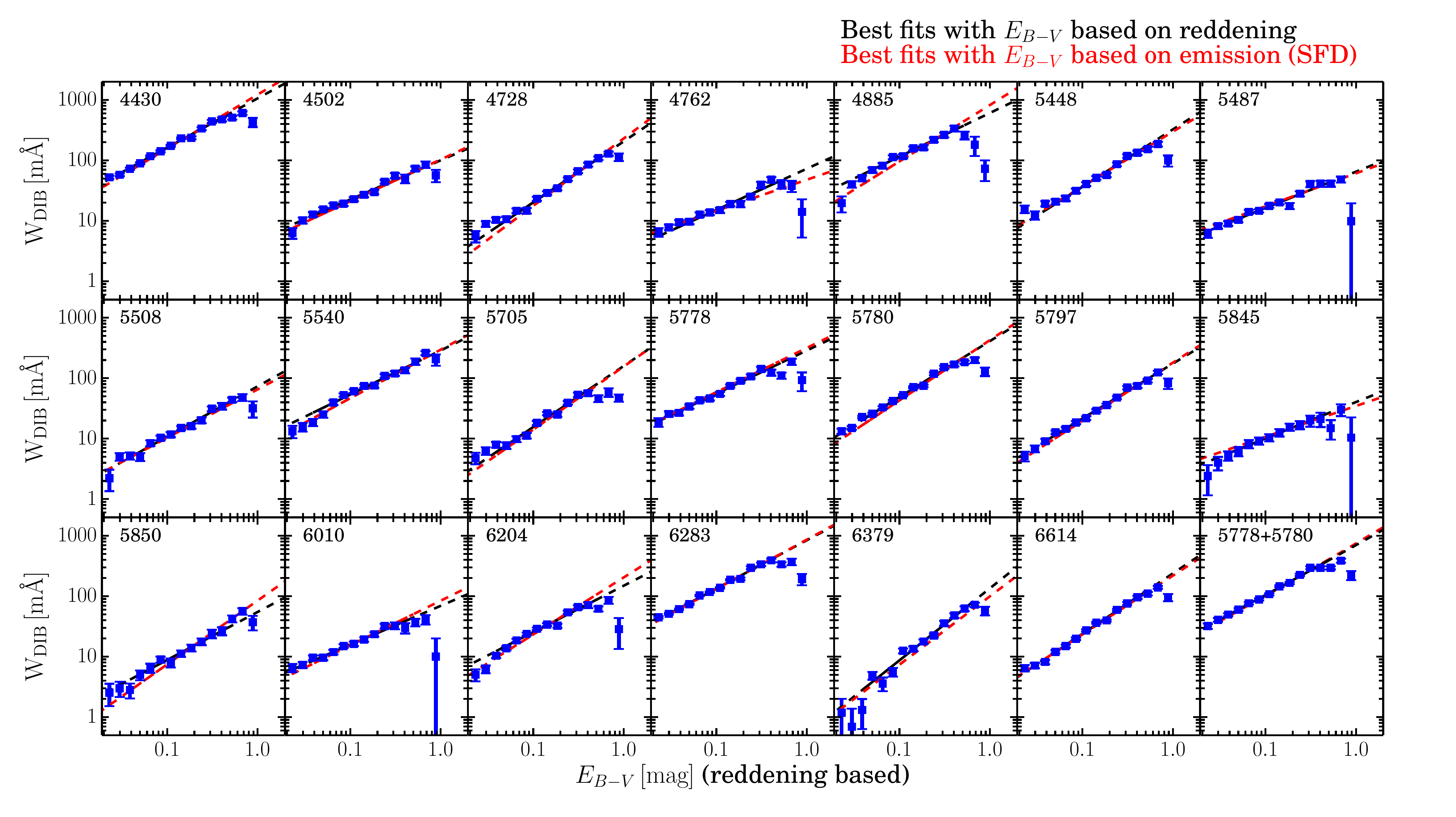}
\caption{Equivalent width measurements of 20 DIBs as a function of $\ebv$ based on reddening estimation. The blue data points show the inverse variance-weighted mean of DIB equivalent widths. The solid black lines are best-fitting power laws derived with reddening-based $\ebv$ from 0.04 to 0.5 and the dashed black lines are extrapolation towards high and low $\ebv$. The dashed red lines are the best-fitting power laws derived with SFD $E_{B-V}$. The two $E_{B-V}$ estimations yield consistent DIB-dust correlations.}
\label{plot:DIB_vs_reddening2}
\end{center}
\end{figure*}

We estimate the Galactic dust reddening of a star by comparing the observed g-r colour of the star to the colour of its reference star and convert into $E_{B-V}$ by using the relation in \citet{Yuan2013}. For individual lines of sight, the noise of $E_{B-V}$ is dominated by the photometric error. To reduce the noise, we estimate the median $E_{B-V}$ for each sky pixel and then combine it with the DIB map shown in Fig.\,\ref{plot:DIB_map}.
Fig.\,\ref{plot:DIB_vs_reddening2} shows the inverse variance-weighted mean of DIB equivalent width as a function of the new reddening-based $\ebv$.
The solid black lines are best-fitting power laws (Table \ref{table:best_fit_power_law2}) with new $E_{B-V}$ measurements with $0.04<E_{B-V}<0.5$ mag and the dashed red lines are the best-fitting power laws with SFD (Fig.~\ref{plot:DIB_vs_reddening}).

We find that the $E_{B-V}$ from SFD tends to be overestimated which can be due to the contamination from background dust or/and systematics in the SFD map \citep[e.g.][]{Arce1999}. Nevertheless, the new derived correlations between DIB absorption strength and $E_{B-V}$ are mostly consistent with the results based on the SFD map. In addition, the different behaviours of DIBs at high Galactic reddening, such as the decline of $\lambda4885$ and the flatten of $\lambda5780$, persist with the absorption-based $E_{B-V}$ estimation, indicating that the different behaviours of DIBs at high Galactic reddening found in the study reflect the intrinsic correlations between DIBs and dust. It is also worth noting that the departure of power laws in the low $E_{B-V}$ regions ($< 0.05$) due to the extragalactic contamination in Fig. 7 has largely reduced by using the reddening based $E_{B-V}$.

\begin{figure*}[!ht]
\begin{center}
\includegraphics[scale=0.45]{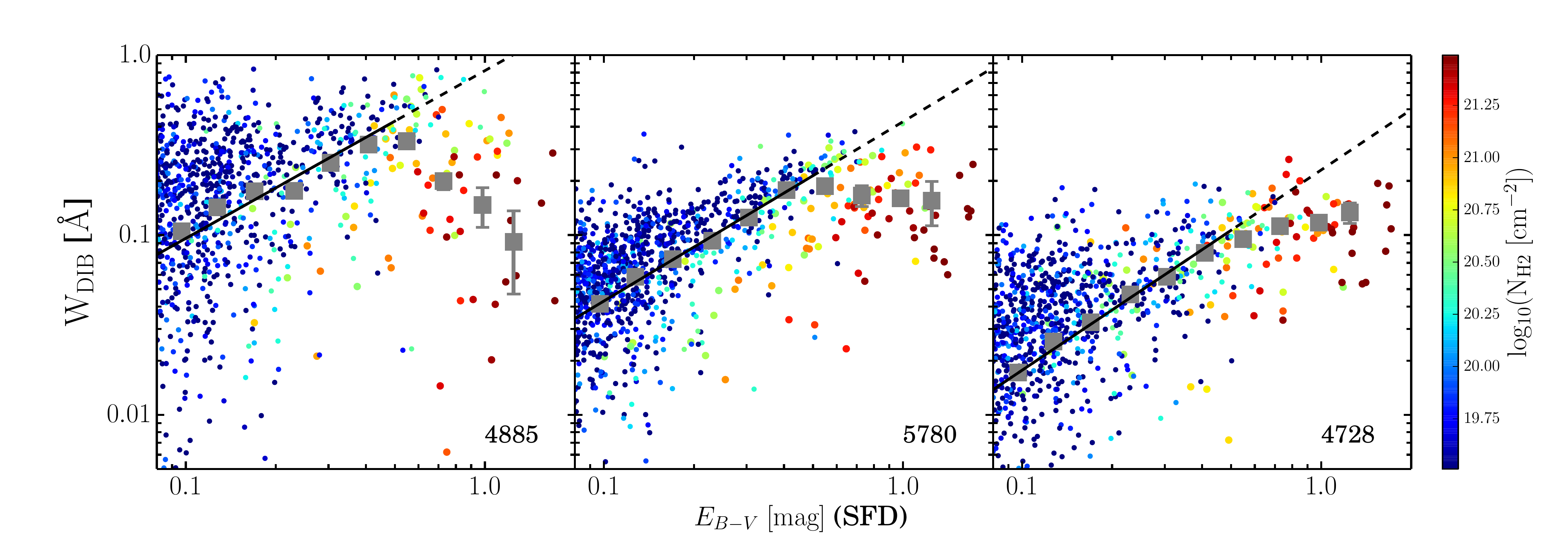}
\includegraphics[scale=0.45]{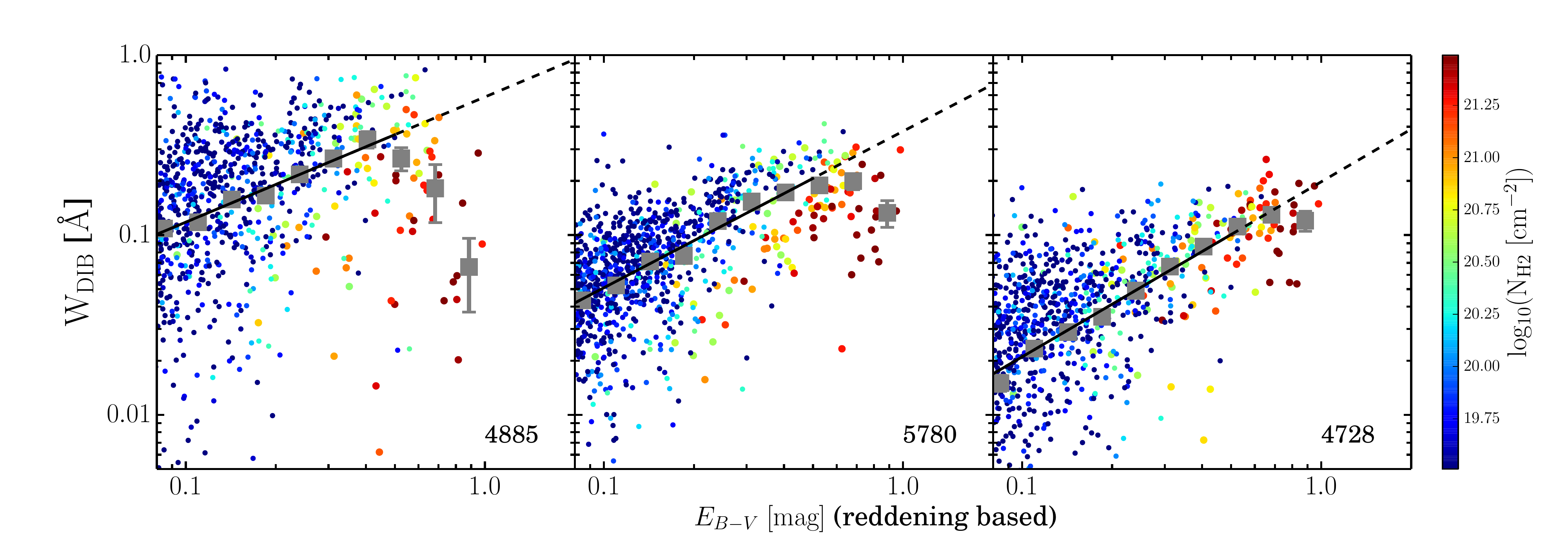}
\caption{Examples of individual DIB equivalent widths as a function of $E_{B-V}$ with $H_{2}$ column densities shown in colour. The grey data points and the black lines are the average values and the best-fitting power laws shown in Fig.~\ref{plot:DIB_vs_reddening} and Fig.~\ref{plot:DIB_vs_reddening2}. \emph{Top}: $E_{B-V}$ from SFD. \emph{Bottom}: $E_{B-V}$ based on reddening. The trends at high $E_{B-V}$ are driven by lines of sight with high $H_{2}$ column densities, which can be explained by the correlations between hydrogen and DIBs shown in Fig.~\ref{fig:main}.}
\label{plot:DIB_vs_reddening3}
\end{center}
\end{figure*}

We note that the decline of some DIBs at high reddening becomes weaker with reddening based $E_{B-V}$ due to the effects on the $E_{B-V}$ from SFD mentioned above. However, in Fig.~\ref{plot:DIB_vs_reddening3}, we present the individual measurements of three DIBs with different trends and show that given a $E_{B-V}$, the equivalent width of DIBs correlates with the hydrogen content along the lines of sight. The colour indicates the $H_{2}$ column densities along the lines of sight. For DIB$\lambda4885$ and DIB$\lambda5780$, the decline and flatten at high $E_{B-V}$ are driven by lines of sight with high $H_{2}$ column densities.
Such trends are found with both SFD and reddening based $E_{B-V}$, indicating that the skin effect is observed. On the other hand, using the reddening based $E_{B-V}$, the equivalent width of DIB$\lambda4728$ can be described by the best-fitting power law from low $E_{B-V}$ to highest $E_{B-V}$ without noticeable deviation. The behaviours of DIBs at high $E_{B-V}$ can be explained by the correlations between DIBs and hydrogen shown in Fig.~\ref{fig:main}: DIB$\lambda4885$ is anti-correlated with molecular hydrogen, DIB$\lambda5780$ has no correlation with molecular hydrogen, and DIB$\lambda4728$ has a positive correlation with molecular hydrogen. 

%----------------------------------
% figure
%----------------------------------
\hspace{-0.5cm}
%----------------------------------

%------------------------------------------------------
\begin{table}[ht] 
\caption{Best-fit parameters characterizing the
     relationships between DIBs and reddening-based \ebv (equation~\ref{eq:w})}
\centering
\begin{tabular}{ccccc}
\hline\hline
$\lambda$ & A & $\gamma$ & ${\rm W/A_{V}}$ \\ [0.5ex] % inserts table 
[\AA] & & & [\AA/mag]\\ [0.5ex]
%heading 
\hline
4430 & $1.06 \pm 0.06$ & $0.82 \pm 0.03$  & 0.42 \\
4502 & $0.10 \pm 0.01$ & $0.65 \pm 0.05$  & 0.05 \\
$\ \, 4728^{*}$ & $0.21 \pm 0.02$ & $1.01 \pm 0.04$  & 0.07 \\
4762 & $0.07 \pm 0.01$ & $0.68 \pm 0.05$  & 0.03 \\
$\ \, 4885^{*}$ & $0.61 \pm 0.06$ & $0.72 \pm 0.05$  & 0.27 \\
$\ \, 5448^{*}$ & $0.32 \pm 0.03$ & $0.95 \pm 0.04$  & 0.11 \\
5487 & $0.06 \pm 0.01$ & $0.60 \pm 0.05$  & 0.03 \\
5508 & $0.07 \pm 0.01$ & $0.83 \pm 0.06$  & 0.03 \\
$\ \, 5540^{*}$ & $0.29 \pm 0.03$ & $0.73 \pm 0.04$  & 0.13 \\
$\ \, 5705^{*}$ & $0.16 \pm 0.02$ & $1.01 \pm 0.05$  & 0.05 \\
$\ \, 5778^{*}$ & $0.29 \pm 0.03$ & $0.71 \pm 0.04$  & 0.13 \\
5780 & $0.41 \pm 0.02$ & $0.93 \pm 0.03$  & 0.14 \\
5797 & $0.17 \pm 0.01$ & $0.91 \pm 0.03$  & 0.06 \\
$\ \, 5845^{*}$ & $0.04 \pm 0.01$ & $0.60 \pm 0.09$  & 0.02 \\
5850 & $0.05 \pm 0.01$ & $0.79 \pm 0.08$  & 0.02 \\
6010 & $0.07 \pm 0.01$ & $0.64 \pm 0.05$  & 0.03 \\
$\ \, 6204^{*}$ & $0.15 \pm 0.01$ & $0.77 \pm 0.04$  & 0.06 \\
$\ \, 6283^{*}$ & $0.84 \pm 0.04$ & $0.80 \pm 0.02$  & 0.34 \\
6379 & $0.13 \pm 0.01$ & $1.19 \pm 0.06$  & 0.03 \\
6614 & $0.24 \pm 0.01$ & $1.00 \pm 0.03$  & 0.08 \\
\hline
\end{tabular}
\begin{tablenotes}
\item *DIBs possibly blended with multiple weak DIBs
\end{tablenotes}
\label{table:best_fit_power_law2}
\vspace{.2cm}
\end{table} 
%------------------------------------------------------

%----------------------------------
% figure
%----------------------------------
\hspace{-0.5cm}
%----------------------------------

\section{DIB-hydrogen correlations with different selections of the sample}
\label{appendix_c}

\begin{figure*}[!ht]
\begin{center}
\includegraphics[scale=0.4]{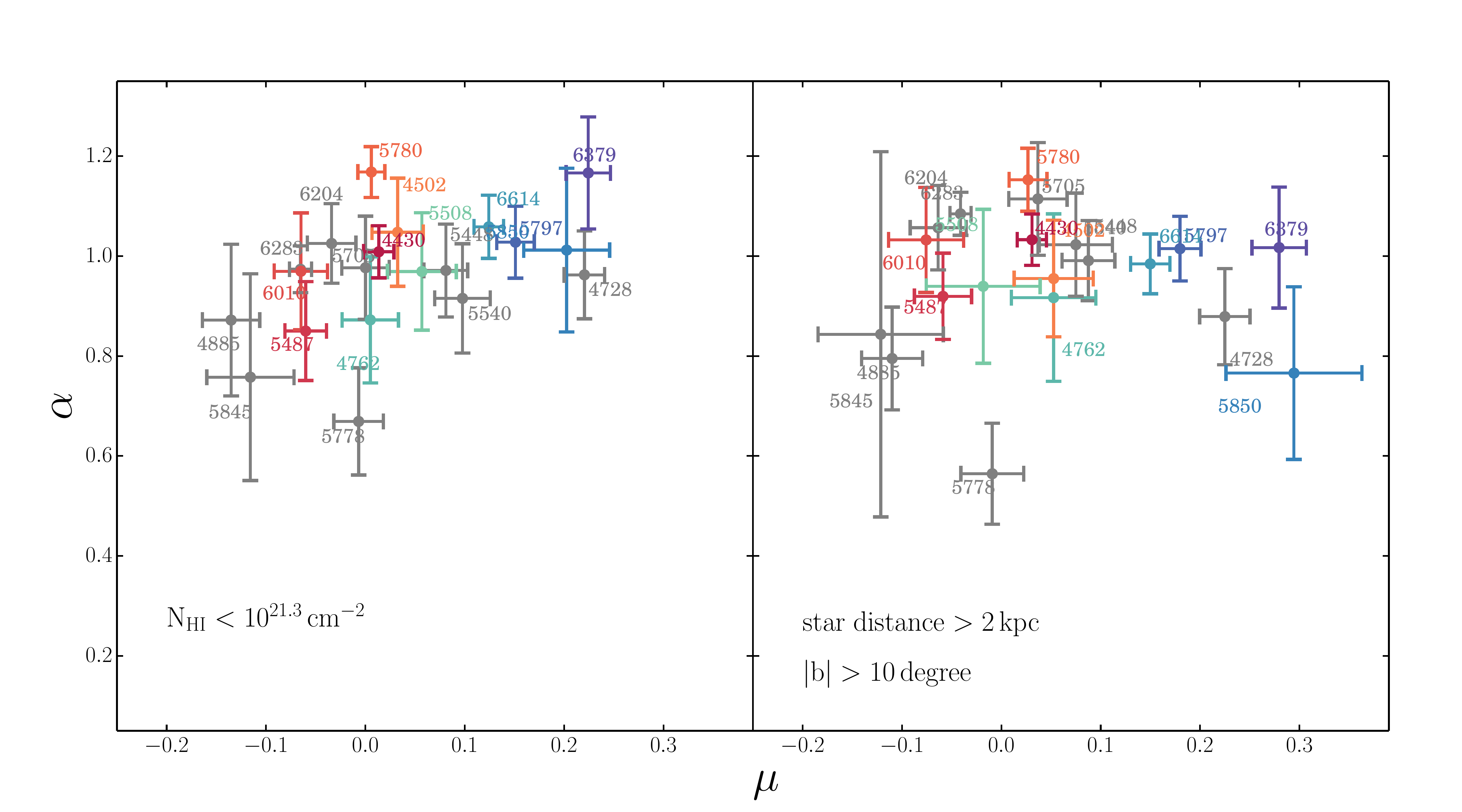}
\caption{Distributions of $\alpha$ and $\mu$ parameters derived from samples with two selections. Two samples are both selected with $E_{B-V}>0.1 \, \rm mag$ and $W_{CO}>0.0 \rm \, K km/s$ and the additional selections are shown in each panel. The left panel shows the same $\alpha$ and $\mu$ distribution as shown in Fig. 12 and the right panel shows the results derived from stars which are two kpc away from the Sun and located above the Galactic disc. The relative positions of DIBs on the $\alpha$ and $\mu$ plane based on two selections are consistent.}
\label{plot:DIB_test}
\end{center}
\end{figure*}

We compare the $\alpha$ and $\mu$ estimation (equation~\ref{eq:main}) using two selections on the data to quantify how the results vary with the selection. The results are shown in Fig.~\ref{plot:DIB_test}. The left panel shows the result with the selection that we applied in the main analysis. The right panel shows the result with a set of stars which are more than 2 kpc away from the Sun. In addition, we also apply a latitude cut to avoid sky regions with high background contamination in $\rm N_{HI}$.  As can be seen, the relative positions of DIBs on the $\alpha$ and $\mu$ plane derived from two selections are consistent with each other. In addition, we have repeated our analysis using DIB$\lambda5780$ as an $\rm N_{HI}$ estimate and find consistent results. Therefore, we conclude the results in the analysis are robust.

%----------------------------------
% figure
%----------------------------------
\hspace{-0.5cm}
%----------------------------------

\end{document}